\pdfoutput=1
\newcommand*{\ATLASLATEXPATH}{atlaslatex-03-00-03/latex/}
\documentclass[cernpreprint,txfonts,UKenglish,texlive=2013]{\ATLASLATEXPATH atlasdoc}

\usepackage{\ATLASLATEXPATH atlaspackage}

\usepackage{\ATLASLATEXPATH atlasbiblatex}

 \usepackage{\ATLASLATEXPATH atlasphysics}

\graphicspath{{atlaslatex-03-00-03/logos/}}

\usepackage{mydocument-defs}
\AtlasTitle{Measurement of charged-particle distributions sensitive to the underlying event
  in $\sqrt{s} = 13$\,TeV proton--proton collisions with the ATLAS detector at the LHC}

\PreprintIdNumber{CERN-EP-2016-28}

\AtlasJournal{JHEP}

\AtlasJournalRef{JHEP 03 (2017) 157}
\AtlasDOI{doi:10.1007/JHEP03(2017)157}

\AtlasAbstract{ %
   We present charged-particle distributions sensitive to the underlying event,
  measured by the ATLAS detector in proton--proton collisions at a
  centre-of-mass energy of 13\,TeV, in low-luminosity Large Hadron Collider
  fills corresponding to an integrated luminosity of \SI{1.6}{\per\nano\barn}.  The
  distributions were constructed using charged particles with absolute
  pseudorapidity less than 2.5 and with transverse momentum greater than
  500\,MeV, in events with at least one such charged particle with transverse
  momentum above 1\,GeV. These distributions characterise the angular
  distribution of energy and particle flows with respect to the charged particle
  with highest transverse momentum, as a function of both that momentum and of
  charged-particle multiplicity. The results have been corrected for detector
  effects and are compared to the predictions of various Monte Carlo event
  generators, experimentally establishing the level of underlying-event activity
  at LHC Run~2 energies and providing inputs for the development of event
  generator modelling. The current models in use for UE modelling typically
  describe this data to 5\% accuracy, compared with data uncertainties of less
  than 1\%. 
}

\begin{document}

\maketitle

\section{Introduction}
\label{sec:intro}
To perform precise Standard Model measurements or to search for new physics
phenomena at hadron colliders, it is important to have a good understanding not
only of the primary short-distance hard scattering process, but also of the accompanying
interactions of the rest of the proton--proton collision -- collectively termed the underlying
event (UE). As the UE is an intrinsic part of the same proton--proton collision
as any ``signal'' partonic interaction, accurate description of its properties
by Monte Carlo (MC) event generators is important for the LHC physics
programme.

In modelling terms, the UE can receive contributions from initial- and
final-state radiation (ISR, FSR), from the QCD evolution of colour connections
between the hard scattering and the beam-proton remnants, and from additional
hard scatters in the same $p$--$p$ collision, termed multiple partonic
interactions (MPI). As it is significantly influenced by physics not currently
calculable from first principles, the measurement of the UE's properties is
crucial not only for better understanding of the mechanisms involved but also to
provide input for empirical tuning of the free parameters of phenomenological UE
models in MC event generators.

It is impossible to uniquely separate the UE from the hard scattering process on
an event-by-event basis, but observables can be defined which are particularly
sensitive to the properties of the UE. Measurements of such observables have
been performed in $pp$ collisions between $\sqrt{s}=\SI{900}{\GeV}$ and \SI{7}{\TeV}
in ATLAS~\cite{Aad:2010fh, Aad:2011qe, Aad:2012jba, Aad:2014hia, Aad:2014jgf},
ALICE~\cite{ALICE:2011ac} and CMS~\cite{Khachatryan:2010pv,
Khachatryan:2015jza,  Chatrchyan:2012tb}. 
UE observables were also previously measured in $p\bar{p}$ collisions in dijet
and Drell--Yan events at CDF, with centre-of-mass energies of
$\sqrt{s} = \SI{1.8}{\TeV}$~\cite{Affolder:2001xt,Acosta:2004wqa} and
\SI{1.96}{\TeV}~\cite{Aaltonen:2010rm}.

In this paper we report the measurement of UE observables with the ATLAS
detector~\cite{Aad:2008zzm} at the LHC, using charged particles in
\SI{1.6}{\per\nano\barn} of proton--proton collisions at a centre-of-mass energy
of \SI{13}{\TeV}.

\section{Underlying-event observables}
\label{sec:observables}
The UE observables in this study are constructed from ``primary'' charged
particles in the pseudorapidity range $\etamod < 2.5$,%
\footnote{ATLAS uses a right-handed coordinate system with its origin at the
  nominal interaction point (IP) in the centre of the detector and the $z$-axis
  along the beam pipe. The $x$-axis points from the IP to the centre of the LHC
  ring, and the $y$-axis points upward. Cylindrical coordinates $(r,\phi)$ are
  used in the transverse plane, $\phi$ being the azimuthal angle around the beam
  pipe. The pseudorapidity is defined in terms of the polar angle $\theta$ as
  $\eta=-\ln\tan(\theta/2)$.} whose transverse momentum (\pt) is required to
be greater than \SI{500}{\MeV}.
Primary charged particles are defined as those with a 
mean lifetime $\tau > \SI{300}{\pico\second}$, which are either directly
produced in $pp$ interactions or from subsequent decays of particles with a
lifetime $\tau < \SI{30}{\pico\second}$. This measurement follows the fiducial
particle definition used in the ATLAS \SI{13}{\TeV} minimum-bias
measurement~\cite{Aad:2016mok} excluding particle species with $\tau$ in the
range \SIrange{30}{300}{\pico\second} and their decay products. The charged
particles falling in this range are strange baryons with a very low
reconstruction probability, whose decays are inconsistently modelled in MC generators.
Their exclusion from the fiducial acceptance definition hence
avoids the need to apply large and poorly defined corrections to particle level
for these species, making the measurement more accurate.

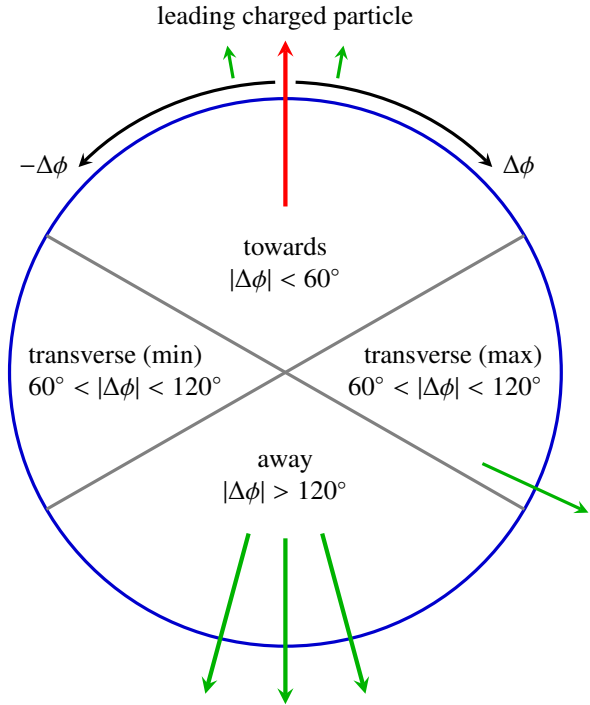
\begin{figure}[tbp]
  \begin{center}
    \begin{tikzpicture}[>=stealth, very thick, scale=1.1]
      \small

      \draw[color=blue!80!black] (0, 0) circle (3.3);
      \draw[rotate= 30, color=gray] (-3.3, 0) -- (3.3, 0);
      \draw[rotate=-30, color=gray] (-3.3, 0) -- (3.3, 0);

      \draw[->, color=black, rotate=-2] (0, 3.5) arc (90:47:3.5) node[right] {$\Delta{\phi}$};
      \draw[->, color=black, rotate=2] (0, 3.5) arc (90:133:3.5) node[left] {$-\Delta{\phi}$};

      \draw[->, color=red, ultra thick] (0, 2) -- (0, 4) node[above,align=center,color=black] {leading~charged~particle};
      \draw[->, color=green!70!black, very thick, rotate around={  10:(0,0)}] (0, 3.6) -- ( 0.0, 4);
      \draw[->, color=green!70!black, very thick, rotate around={ -10:(0,0)}] (0, 3.6) -- ( 0.0, 4);
      \draw[->, color=green!70!black, ultra thick] (0, -2) -- ( 0.0, -4);
      \draw[->, color=green!70!black, ultra thick, rotate around={-15:(0,-0.3)}] (0, -2) -- ( 0.0, -4);
      \draw[->, color=green!70!black, ultra thick, rotate around={ 15:(0,-0.3)}] (0, -2) -- ( 0.0, -4);
      \draw[->, color=green!70!black, very thick, rotate around={ 65:(0,0)}] (0, -2.6) -- ( 0.0, -4); 

      \draw (0,  1.3) node[align=center] {towards\\ $|\Delta\phi| < 60^\circ$};
      \draw (0, -1.3) node[align=center] {away\\ $|\Delta\phi| > 120^\circ$};
      \draw (3.2, 0) node[align=right, anchor=east] {transverse~(max)\\ $60^\circ < |\Delta\phi| < 120^\circ$};
      \draw (-3.2, 0) node[align=left, anchor=west] {transverse~(min)\\ $60^\circ < |\Delta\phi| < 120^\circ$};
    \end{tikzpicture}
    \caption[Definition of UE regions]{Definition of regions in the azimuthal
      angle with respect to the leading (highest-\pt) charged particle, with arrows representing
      particles associated with the hard scattering process and the leading
      charged particle highlighted in red.  Conceptually, the presence of a
      hard-scatter particle on the right-hand side of the transverse region,
      increasing its \ptsum, typically leads to that side being identified as
      the ``trans-max'' and hence the left-hand side as the ``trans-min'', with
      maximum sensitivity to the UE.}
    \label{fig:ueregions}
  \end{center}
\end{figure}

This measurement uses the established form of UE observables~\cite{Aad:2010fh, Aad:2011qe,
  Aad:2012jba, Aad:2014hia, Aad:2014jgf, ALICE:2011ac, Khachatryan:2010pv,
  Chatrchyan:2012tb, Affolder:2001xt, Aaltonen:2010rm}, in which the
azimuthal plane of the event is segmented into several distinct regions with
differing sensitivities to the UE. As illustrated in Figure~\ref{fig:ueregions},
the azimuthal angular difference with respect to the leading (highest-\pt) charged particle,
$|\Delta\phi|=|\phi-\phi_\text{lead}|$, is used to define the regions:
\begin{itemize}
\item $|\Delta\phi| < 60\degr$, the ``towards region'';
\item $60\degr < |\Delta\phi| < 120\degr$, the ``transverse region''; and
\item $|\Delta\phi| > 120\degr$, the ``away region''.
\end{itemize}
As the scale of the hard scattering increases, the leading charged particle acts
as a convenient indicator of the main flow of hard-process energy. The towards
and away regions are
dominated by particle production from the hard process and are hence relatively
insensitive to the softer UE. In contrast, the transverse region is more
sensitive to the UE, and observables defined inside it are the primary focus of
UE measurements.

A further refinement is to distinguish on a per-event basis between the more and
the less active sides of the transverse region~\cite{Marchesini:1988hj,
  Pumplin:1997ix}, defined in terms of their relative scalar sums of primary
charged-particle \pt and termed ``trans-max'' and ``trans-min'' respectively. %
The trans-min region is relatively insensitive to wide-angle emissions from the
hard process, and the difference between trans-max and trans-min observables
(termed the ``trans-diff'') hence represents the effects of hard-process
contamination. In this analysis, an event must have a non-zero primary
charged-particle multiplicity in the trans-min region in order to be included in
either the trans-min, -max, or -diff observables.

The variables measured in this analysis, constructed using charged particles
with $\pt > \SI{0.5}{\GeV}$ and $\etamod < 2.5$, with a higher-\pT requirement
of $\ptlead > \SI{1}{\GeV}$ 
placed on the leading charged particle, are described in Table~\ref{tab:obs}.
These variables are divided into two groups: first the ``binned'' per-event or
per-particle quantities used to define the horizontal axes in the unfolded
observables of Section~\ref{sec:result}; and secondly the 
``averaged'' mean values of distributions of per-event quantities to be studied
as functions of the binned variables -- a construction known as a
``profile''. The second-group variables are defined for each
bin 
and (except for \meanmeanpt, in which the \etaphi area factors cancel) are
scaled by the corresponding \detadphi areas to give densities comparable between
all such measurements, including between various experiments and collider
energies.
The area factor $\delta\phi$ is \num{2\pi/3} for the toward, transverse \& away
regions, whereas it is \num{\pi/3} for the single-sided trans-min \& trans-max
regions, and $2\pi/n_\mathrm{bins}$ for each of the $n_\mathrm{bins}$ equally
sized bins in distributions plotted against \Adphi.  Due to the $\etamod < 2.5$
fiducial acceptance and the $\eta$-independence of the region definitions,
$\delta\eta = 5$ in all cases. The profile observables are implemented as
profile histograms, presenting the 
mean values of the ``averaged'' variables as measured in each bin of another
observable. These hence measure the degree of correlation between two event
features, either between the UE and hard scattering, or between different UE
aspects.  The mean charged-particle momentum \ptmean is constructed on an
event-by-event basis and then averaged over all events to give \meanmeanpt.

The majority of underlying-event observables study the dependences of the
averaged quantities on the transverse momentum of the leading object -- here the
leading charged particle. The development of this 
from low to high \pt corresponds to 
the smooth transition from 
``minimum bias'' interactions 
to the hard-scattering regime focused on by most LHC analyses, and the
correlation distributions characterise how soft QCD effects co-evolve with the
hard process through this transition. This analysis also studies the dependence
of the observables on the azimuthal angle with respect to the leading particle
and each region's charged-particle multiplicity.  For the observables studied as
a function of relative azimuthal angle, the leading particle is excluded from
the spectrum.

\begin{table}[tbp]
  \begin{center}
    \caption[Observables]{Definitions of the measured observables in terms of primary charged
      particles. The upper group of observables are used to define the $x$-axes of the plots
      to be shown in Section~\ref{sec:result}, and the lower group are the mean values of
      distributions constructed in each $x$-bin and plotted on the $y$-axes. The \detadphi
      scale factors convert the raw measurements of regional \nch and \sumpt into densities
      per unit \etaphi and their values change depending on the region/bin-sizes being
      considered, so that the density variables are everywhere directly comparable.}
    \label{tab:obs}
    \vspace{1ex}
    \begin{tabular}{lll}
      \toprule
      Symbol                                         & Description \\
      \midrule
      \multicolumn{2}{l}{\textsc{Binned variables}}\\
      \addlinespace
      ~\ptlead                                        & Transverse momentum of the leading charged particle \\
      \addlinespace
      ~\nchtrans                                      & Number of charged particles in the transverse region \\
      \addlinespace
      ~\Adphi                                         & Absolute difference in particle azimuthal angle from the leading charged particle \\
      \addlinespace
      \addlinespace
      \multicolumn{2}{l}{\textsc{Averaged variables}}\\
      \addlinespace
      ~$\left\langle \nchgdetadphi \right\rangle$     & Mean number of charged particles per unit \etaphi (in radians) \\
      \addlinespace
      ~$\left\langle \ptsumdetadphi \right\rangle$    & Mean scalar \pt sum of charged particles per unit \etaphi (in radians) \\
      \addlinespace
      ~$\left\langle \meanpt \right\rangle$           & Mean per-event average \pt of charged particles ($\ge 1$~charged particle required) \\
      \bottomrule
    \end{tabular}
  \end{center}
\end{table}

\section{Monte Carlo modelling of the underlying event}
\label{sec:mc}
As a physics process related to the bulk structure of protons and not calculable
from first-principles perturbative QCD, the underlying event is modelled in
Monte Carlo event generator programs by various phenomenological approaches.
The scattering subprocess type which contributes most to UE observables is
non-diffractive inelastic scattering. In event generator implementations,
model-specific choices are made to regulate these processes' QCD divergences at
low scales. In this analysis the observable definitions restrict the effect of
diffractive scattering, i.e.~colour-singlet exchange, to play a minor role: the
\ptmean profile observables have been found to be completely insensitive to
diffraction, while the \nch and \ptsum profiles exhibit a 2\% effect at low \ptlead
and the whole of the \nch vs.~\dphi distribution is affected by diffraction at a
1--2\% level.

This section reviews relevant features of the
\pythiaeight~\cite{Sjostrand:2007gs,Sjostrand:2006za},
\herwigseven~\cite{Bellm:2015jjp} and \epos~\cite{Pierog:2009zt,Pierog:2013ria}
MC event generator models, which are used in this study either for
data 
correction or for comparison to the final corrected data distributions. A
summary of the Monte Carlo generator configurations used is given in
Table~\ref{tab:mcTunes}.

\begin{description}
\item [\pythiaeight.18\,/\,8.21:] The \pythia generator family is very widely
  used at the LHC and elsewhere for event modelling, and \pythiaeight is its
  most recent 
  release series. Its key features are leading-logarithmic initial- and
  final-state parton showers and a Lund string hadronisation model, in addition
  to particle decays and soft-QCD modelling. Hard-process calculations are
  performed either via an internal set of leading-order matrix elements, or by
  reading externally generated hard scattering events.

  The \pythiaeight approach to soft-QCD modelling uses a parameterised
  total $pp$ cross-section, which is split into elastic, diffractive, and
  non-diffractive (ND) inelastic scattering subprocesses. The first two of these
  again have parameterised cross-sections, and the hadronic ND inelastic
  cross-section is set by subtracting their contributions from the total.

  The ND contribution is modelled using perturbative $2 \to 2$ QCD matrix
  elements, dominated by $t$-channel gluon exchange. As the partonic
  cross-section for this exceeds the hadronic ND cross-section at low \pt,
  the existence of MPI 
  is implied and may be used to regularise the cross-section
  growth. \pythiaeight uses an evolved form of the 
  Sj\"ostrand--van~Zijl MPI model~\cite{Sjostrand:1987su} in which the eikonal
  formalism is applied to give a model with a Poisson distribution of multiple
  perturbative scatterings whose mean rate depends on the scale of the hard
  process (interpreted as the reciprocal of the $pp$ impact parameter), the
  proton form factor, and the ratio of hadronic to partonic ND cross-sections.
  An ansatz is used to regularise cross-section growth at low \pt, with a weak
  power-law evolution of the regularisation parameter with $\sqrt{s}$.  The
  current form of the model also interleaves MPI emissions with parton shower
  evolution, allows several forms of matter overlap parameterisation, and
  interacts with a non-perturbative annealing procedure for reconfiguration of
  colour strings during hadronisation (the ``colour reconnection'' mechanism).

  As MPI and hadronisation modelling are phenomenological and even the
  perturbative parton shower formalism has some configurational freedom,
  many parameter optimisations (``tunes'') of \pythiaeight have been
  performed. The following are considered in this study:

  \begin{itemize}
  \item ATLAS's dedicated underlying-event tune is
    ``A14''~\cite{ATL-PHYS-PUB-2014-021}. Its configuration is based on the
    NNPDF2.3\,LO~\cite{Ball:2012cx} parton density function (PDF), and was
    optimised for the description of several underlying event and jet radiation
    observables, including jet, Drell--Yan, and $t\bar{t}$ data, with an
    emphasis on high-scale events.

  \item An alternative tune, ``Monash''~\cite{Skands:2014pea}, is used for
    comparison. Like A14, it was constructed using Drell--Yan and underlying-event
    data from ATLAS, but also data from CMS, from the SPS, and from the Tevatron
    in order to constrain energy scaling. It also uses the NNPDF\,2.3\,LO PDF,
    and has more of a general-purpose / low-\pt emphasis than A14. This tune
    gives an excellent description of the ATLAS 7, 8 and \SI{13}{\TeV}
    minimum-bias \pt spectra~\cite{Aad:2010ac,Aad:2016xww,Aad:2016mok}.

  \item The ATLAS minimum-bias tune ``A2''~\cite{ATL-PHYS-PUB-2011-014} was used for
    deriving detector corrections. This is based on the MSTW2008\,LO
    PDF~\cite{Martin:2009iq}, and was tuned using ATLAS minimum-bias data at
    \SI{7}{\TeV} for the MPI parameters, in addition to the older \pythiaeight tune
    ``4C'' values for fragmentation parameters. It provides a good description of
    minimum-bias and transverse energy flow data~\cite{Aad:2012mfa}.

  \end{itemize}

  The \pythiaeight predictions shown in this paper use large MC samples from
  versions 8.185 and 8.186, but checks against the newer 8.2xx release series
  (specifically, version 8.210) found no distinguishable difference.

\begin{table}[tbp]
   \begin{center}
     \caption{Details of the MC models used. Some tunes are
      focused on describing the minimum-bias (MB) distributions better, while
      the rest are tuned to describe underlying event (UE) or double parton
      scattering (DPS) distributions.} 
    \label{tab:mcTunes}
    \vspace{1ex}
    \begin{tabular}{llllll}
      \toprule
      Generator     & Version & Tune  & PDF &  Focus  & From  \\
      \midrule
      \pythiaeight  & 8.185 & A2      & MSTW2008\,LO & MB     & ATLAS   \\
      \pythiaeight  & 8.185 & A14     & NNPDF2.3\,LO & UE     & ATLAS   \\
      \pythiaeight  & 8.186 & Monash  & NNPDF2.3\,LO & MB/UE  & Authors \\
      \herwigseven  & 7.0.1 & UE-MMHT & MMHT2014\,LO & UE/DPS & Authors \\
      \epos         & 3.4   & LHC     & ---          & MB     & Authors \\
      \bottomrule
    \end{tabular}
  \end{center}
\end{table}

\item [\herwigseven.0.1:] The \herwig family of MC
  generators has also been heavily used in many collider physics studies, and
  \herwigseven is the most recent major series~\cite{Bahr:2008pv,Bellm:2015jjp}.
  Like \pythia, it is
  a fully exclusive hadron-level generator, containing leading-logarithmic
  parton showers, hadronisation and decays, an MPI mechanism, and capabilities
  for parton-shower matching to higher-order hard processes. It uses a cluster
  hadronisation scheme, and angular-ordered and dipole parton showers.

  The soft-QCD modelling in \herwigseven (and \herwigpp before it) uses an
  eikonal model similar to \pythia's, but with some distinctions. The same
  treatment with Poisson-distributed simulation of many independent perturbative
  QCD scatters is then used, but with a simpler MPI parameterisation than in
  \pythia: the functional form of the proton electromagnetic form factor is used
  to parameterise the hadronic matter distribution rather than \pythia's very
  flexible parameterisation, and a continuation of eikonal scattering down to
  very low \pt values is applied in place of \pythia's phenomenological
  regularisation of eikonal scattering below the \pt cutoff. The \herwig soft
  eikonal scattering model uses distinct matter distributions for soft and hard
  MPI scattering, and introduces a $\sqrt{s}$-dependence of MPI parameters
  similar to that found in \pythia. The parameters of this model are highly
  constrained by fits to total cross-section and elastic scattering
  data~\cite{Bahr:2008pv,Bahr:2008wk}.
  A colour-disruption mechanism is used in hadronisation, as a cluster-oriented
  analogy to the \pythia colour reconnection, to improve the quality of
  minimum-bias observable description.

  The \herwigseven default tune, H7-UE-MMHT with the MMHT2014\,LO
  PDF~\cite{Harland-Lang:2014zoa}, has been used.
  This tune, like its \Herwigpp predecessors, is based on LHC and
  Tevatron underlying-event measurements, as well as
  double parton scattering data~\cite{Seymour:2013qka}. It provides a good
  description of all these observables for $\sqrt{s}$ from Tevatron
  \SI{300}{\GeV} to LHC \SI{7}{\TeV}.

\item [\epos\,3.4:] An alternative approach is taken by \epos, a specialist
  soft-QCD/cosmic-ray air-shower MC generator based on an implementation of
  parton-based Gribov--Regge theory~\cite{Pierog:2009zt}. This is a QCD-inspired
  effective-field theory describing the hard and soft scattering
  simultaneously. It incorporates elements of collective flow modelling from
  nuclear and heavy-ion physics, using hydrodynamic flow modelling in
  high-density regions and a string-based hadronisation model elsewhere. As a
  result the \epos calculations do not make use of standard parton density
  functions. Using this version, equivalent to the so-called ``LHC tune'' of
  version 1.99~\cite{Pierog:2013ria}, \epos gives a very good description of
  ATLAS's \SI{13}{\TeV} minimum-bias data, including the tails of distributions
  where UE physics should be involved, but as it lacks a dedicated
  hard scattering component it is unclear how accurate its description of UE
  correlations can be.

\end{description}

\subsection*{Detector-level simulation}

The \pythiaeight~A2 sample and an MC simulation of single particles distributed
to populate the high-\pt region were used to derive the detector corrections for
these measurements.  Smaller samples produced with the \epos generator were used
to cross-check the validity of the correction procedure.

These events were processed through the ATLAS detector simulation
framework~\cite{Aad:2010ah}, which is based on
\textsc{Geant\,4}~\cite{Agostinelli:2002hh}. They were then reconstructed and
analysed using the same processing chain as for data. In all Monte Carlo samples
the distribution of the primary vertex position was reweighted to match the
distribution in data.

\section{The ATLAS detector}
\label{sec:detector}
The ATLAS detector is described in detail in Ref.~\cite{Aad:2008zzm}. In this
analysis, the trigger system and inner tracking detectors are of particular relevance.

The inner tracking detector is immersed in the \SI{2}{\tesla} axial magnetic
field of a superconducting solenoid, and measures the trajectories of charged
particles in the pseudorapidity range $|\eta| < 2.5$ with full azimuthal
coverage.  It consists of a silicon pixel detector (pixel), a silicon microstrip
detector (SCT) and a straw-tube transition radiation tracker (TRT), split into a
barrel and two endcap components.  The barrel consists of 4 pixel layers, 4
layers of SCT modules with back-to-back silicon strip sensors, and 73 layers of
TRT straws; each endcap has 3 pixel layers, 9 SCT layers, and 160 TRT straw
layers.  The pixel, SCT and TRT have $r$--$\phi$ position resolutions of
\SI{10}{\micro\metre}, \SI{17}{\micro\metre}, and \SI{130}{\micro\metre}
respectively, and the pixel (apart from the innermost barrel layer) and SCT have
resolutions of \SI{115}{\micro\metre} and \SI{580}{\micro\metre} respectively in
the $z$-direction for the barrel modules and in the $r$-direction for the
endcaps.

The innermost pixel layer, the ``insertable B-layer''~(IBL), 
was added between LHC Runs~1 and~2, around a new narrower (radius of
\SI{25}{\milli\metre}) and thinner beam pipe~\cite{Capeans:1291633}. It is
composed of 14 lightweight staves arranged in a cylindrical geometry, each made
of 12 silicon planar sensors in its central region and $2 \times 4$
three-dimensional sensors at the ends.
The IBL pixel dimensions are $50 \times \SI{250}{\micro\metre}$ 
in the $\phi$ and $z$ directions (compared with
$50 \times \SI{400}{\micro\metre}$ for other pixel layers).  The smaller radius
and the reduced pixel size result in improvements of both the transverse and
longitudinal impact parameter resolutions.  In addition, the services for the
existing pixel detector were upgraded, significantly reducing the material in
$|\eta| > 1.5$, in particular at the boundaries of the active tracking volume.
A track from a charged particle traversing the barrel detector typically has
12~silicon measurement points (hits), of which 4 are pixel and 8 are SCT, and
more than 30~TRT straw hits.

The ATLAS detector has a two-level trigger system: level~1 and the high-level
trigger~\cite{ATL-DAQ-PUB-2016-001}.  Events used in this analysis were required
to satisfy level-1 triggers using the 
minimum-bias trigger scintillators (MBTS). 
These are mounted at each end of the detector in front of the liquid-argon
endcap-calorimeter cryostats at $z = \pm \SI{3.56}{\metre}$ and are segmented
into two rings in pseudorapidity ($2.07 < |\eta| < 2.76$ and
$2.76 < |\eta| < 3.86$), with 8 azimuthal sectors in the inner ring and 4 in the
outer. The MBTS scintillator was replaced for Run~2 due to radiation damage
incurred in Run~1. The MBTS triggers fired if at least one MBTS hit from either
side of the detector was recorded above threshold.

\section{Event and object selection}
\label{sec:selection}

This measurement uses data taken in a special configuration of the LHC, with low beam
currents and reduced beam focusing, producing a low mean number of interactions
per bunch-crossing, \avgmu, between 0.003 and 0.03.  This configuration and
event selection were earlier used and documented in detail in the \ATLAS
\SI{13}{\TeV} minimum-bias analysis~\cite{Aad:2016mok}, and so only
a summary is given here.

\paragraph{Trigger:}
Events were selected from colliding proton bunches using a trigger which
required one or more MBTS counters above threshold on either side of the
detector. The efficiency of this trigger was measured
to be 99\% for low-multiplicity events and to rapidly rise to 100\% for events with higher track
multiplicities. The trigger requirement does not bias the \pt and $\eta$
distributions of selected tracks in this analysis due to the \leadpt >
\SI{1}{\GeV} requirement.

\paragraph{Primary vertex:}
Each event was required to contain a primary vertex 
reconstructed from at least two tracks with $\pt > \SI{100}{\MeV}$ and selection
requirements specific to vertexing~\cite{Piacquadio:2008zzb}. The canonical
primary vertex was identified as that with the highest $\sum \pt^2$ of its
associated tracks. To reduce contamination from events with more than one
interaction in a bunch crossing (``pile-up''), events containing more than one
primary vertex with four or more associated tracks were removed.  The
contributions from non-collision backgrounds and 
events where two interactions are reconstructed as a single vertex were studied
in data and found to be negligible.

A total of 66~million data events passed the trigger and vertex selection
requirements for this analysis.

\vspace*{2mm}
\paragraph{Tracks:}
Tracks were reconstructed 
starting from hits in the silicon detectors and then extrapolated to include
information from the~TRT. Each track required hits in both the pixel system and
the SCT, including a requirement of a hit in the innermost expected pixel
layer to reject secondary particles.

All tracks were reconstructed within the geometric acceptance of the inner detector,
$|\eta| < 2.5$. In this analysis, all selected tracks were additionally required
to have a transverse momentum above \SI{500}{\MeV}, and both the transverse
impact parameter and the projected longitudinal impact parameter %
with respect to the primary vertex were required to be less than
\SI{1.5}{\milli\metre}.  For high-\pt tracks, a further requirement was placed
on the $\chi^2$ probability of the track fit, to suppress mismeasured tracks.
The selected events were additionally required to contain at least one track
with a transverse momentum above \SI{1}{\GeV}.

\paragraph{Backgrounds:}
The contributions from non-collision background events, events with more than
one interaction, and fake tracks (those formed by a random combination of hits
or from a combination of hits from several particles) were found to be
negligible. The contribution from secondary particles was estimated as in
Ref.~\cite{Aad:2016mok}.

\section{Correction to particle level}
\label{sec:correction}
In order to compare the measured underlying-event distributions with model
predictions in the particle-level fiducial phase space, the observables have been
corrected for detector effects.
These corrections include explicit accounting for inefficiencies due to the
trigger selection, vertexing, and track reconstruction,
by weighting the events and tracks by inverse efficiencies.
A further correction has been applied to account for non-linear effects,
particularly azimuthal re-orientation of the event which occurs should the leading charged
particle not be reconstructed. In this situation the identification of the
towards, transverse, and away regions can differ from that at particle level,
leading to ``wrong'' track--region associations. Re-orientation mainly affects
events with a low number of charged particles, and has been corrected using a
dedicated method.
The track-weight correction is described first, followed by the method to
account for event re-orientation.

\subsection{Event and track weighting}

The weighting procedure for correcting measured distributions to particle level
is affected by the primary vertex reconstruction efficiency, the tracking efficiency, the rate
of non-primary particles, and the rate of charged strange baryons. Before
defining the weights, we summarise these aspects of the vertexing and tracking
performance:

\paragraph{Vertexing efficiency:}

The vertex reconstruction efficiency, \vtxeff, was determined from data by
taking the ratio of the number of selected events with a reconstructed vertex to
the total number of events with the requirement of a primary vertex removed.
The efficiency was close to 100\% except for events in which exactly one
$\pt > \SI{500}{\MeV}$ analysis track was found in the inner detector -- for
these the vertexing efficiency was found to be 90\%. Since the leading-track
$\pt > \SI{1}{\GeV}$ requirement in the analysis is correlated
with 
higher track multiplicities, the effect of the vertex reconstruction on the
analysis is small.

\paragraph{Tracking efficiency:}

The efficiency to reconstruct primary charged particles was determined from
simulation by matching MC-generator primary charged particles to tracks
reconstructed from simulated inner-detector hits. For particles at the analysis
transverse-momentum cutoff of $\pt = \SI{500}{\MeV}$, the track
reconstruction is $\sim\!85\%$ efficient for central pseudorapidity, and
decreases to 65\% for $\etamod \sim 2.5$ at the edge of the inner tracking detector; these
efficiencies are higher for tracks with higher \pt, for example 90\% and 80\%
respectively for $\pt = \SI{10}{\GeV}$.

Most of the loss in tracking efficiency is due to particle interactions
with the detector material and support structures, and hence a good description
of the detector material is needed. The ATLAS inner tracker upgrades for LHC Run~2
necessitated a new study of the detector material uncertainties, which are
detailed in Ref.~\cite{Aad:2016mok}. The track reconstruction efficiency for
$|\eta| > 1.5$ is corrected using a method that compares the efficiency to extend
a track reconstructed in the pixel detector into the SCT in data and simulation.

\paragraph{Rate of non-primary tracks:}
There are several sources of non-primary tracks: 1)~tracks from hadronic
interaction of particles with the detector material,\; 2)~decay products from
particles with strange quark content, mostly $K^0$ and $\Lambda^0$ decays,\;
3)~photon conversions. The last of these is negligible for the tracks with
$\pt > \SI{500}{\MeV}$ used in this analysis.
The rate of non-primary tracks was determined by comparing side-band fits of MC
simulation impact parameter distributions to the data, leading to an estimated
2.3\% of non-primary tracks in data for the $\pt > \SI{500}{\MeV}$ track selection,
and a smaller effect for higher-\pt tracks.

\paragraph{Fraction of charged strange baryons:}
Charged strange baryons and their decay products have low reconstruction
efficiency, unless they have large transverse momentum. 
With the fiducial primary particle definition used here, the fraction of tracks due to strange baryons with $\pt = \SI{10}{\GeV}$ is 1\%,
while it is much smaller than 1\% at lower \pt. With our fiducial primary
particle definition, tracks originating from charged strange baryons are
classified as a background. An estimate of their contribution was made using the
\epos generator, a choice motivated both by \epos providing the best current
description of charged strange baryon rates as measured by the ALICE
experiment~\cite{Aamodt:2011zza}, and by substantial variation of strange baryon
modelling between different MC generators.

\paragraph{Weighting:}
The effect of events lost due to the trigger and vertex requirements was
corrected by an event-by-event weight,
\begin{equation}
\evw\! \left( \nselbs, \eta \right) =  \frac{1}{\trigeff\! \left(\nselbs\right)} \cdot \frac{1}{\vtxeff\! \left(\nselbs,\eta\right)} \, ,
\label{eq:corr:wev}
\end{equation}
where \trigeff is the trigger efficiency and $\nselbs$ is the multiplicity of
``beam line'' selected
tracks, 
which have the same selection requirements as analysis tracks except that no
vertexing requirement is applied and and hence there is no restriction on the
longitudinal impact parameter. 

To correct for inefficiencies in the track reconstruction, the distributions of
the selected analysis tracks were corrected with track-by-track weights,
\begin{equation}
\label{eq:corr:wtrk}
   w_\mathrm {trk}(\pt, \eta) =  \frac{1}{\trkeff\left( \pt, \eta \right)}
   \cdot \left(1 - \fsec(\pt, \eta) - \fokr(\pt , \eta) - \fsb(\pt) \right) \, ,
\end{equation}
where \trkeff is the tracking efficiency, and \fsec, \fokr and \fsb are
respectively the fractions of non-primary tracks, of out-of-kinematic-range
tracks (i.e. tracks mis-reconstructed as within the fiducial $\pt$ and $\eta$ acceptances, but which actually originated from outside those ranges), and of weakly decaying charged strange baryons.

This track weight was used in the construction of
the \nch, \sumpt, and \dphi distributions, and the mean~\pt in the event,
determined using the weighted average
$\ptmean = \sum_{i\in\,\mathrm{tracks}} p^{\,i}_\mathrm{T} w^{\,i}_\mathrm{trk}
/ \sum_{i\in\,\mathrm{tracks}} w^{\,i}_\mathrm{trk}$.

For the distributions binned in charged multiplicity, this weight was not used
to correct the binning variable, since this induces migrations subject to
fluctuations not accounted for by track weighting. This correction was instead
handled by the residual correction method described in the next section.

\subsection{Re-orientation correction}

The re-orientation correction was based on the HBOM method~\cite{Monk:2011pg},
in which the effect of the detector and reconstruction algorithms on an
observable (i.e. a histogram bin value) is treated as an operator $A$. The
observed reconstruction-level value of an observable, $X^\text{obs}$, is hence
related to its true value $X^\text{true}$ as $X^\text{obs} = A X^\text{true}$,
but one can continue to apply $A$ to the modified event, so that the $k$th additional
application gives $X^\text{obs}_k = A^k X^\text{obs} = A^{k+1} X^\text{true}$. If the evolution of
the observable's value is a smoothly varying function under such iterations, an
$n^\text{th}$-order polynomial function $P_{\!n}(k)$ can be fitted to
$X^\text{obs}_k$ for $k \ge 0$, and then extrapolated back to $k = -1$,
i.e.~$X^\text{true}$.
This procedure is carried out independently for each bin of each distribution.
It is distinct from the unfolding approach used in
Refs.~\cite{Aad:2014hia} and~\cite{Aad:2016mok}, chosen because it is fully
data-driven, i.e. does not rely on simulation performance or require reweighting of MC
events to data.

In this analysis, $A$ encodes the effects of track reconstruction, i.e.~each
application of $A$ in principle smears the track kinematics and considers the
possibility that some tracks would not have been reconstructed -- directly inducing
re-orientation should the leading track experience such a fate. These effects are
described by the measured tracking resolutions and efficiencies, as functions of
track $\eta$ and \pt, but the efficiency was found to dominate and the
resolution to be negligible, so $A$ reduces to random discarding of tracks
according to their reconstruction efficiency.
For the profile distributions, a weight proportional to the inverse of tracking
efficiency (as above) was applied in each HBOM iteration step, $k$, to make the
effect on the observable less dramatic and more easily parameterisable. 

To avoid correlations between observables with different $k$ values, each
$X^\text{obs}_k$ was calculated independently, using a distinct random seed and
$k$ iterations of track discarding starting from the set of reconstructed
($k=0$) tracks. Second-order polynomials and a maximum of $k=4$ HBOM iterations
were used as the nominal 
configuration, with variations 
being used to derive an extra systematic uncertainty, described in the following
section.

\section{Systematic uncertainties}
\label{sec:systematics}
\vspace{-1ex}

Several categories of systematic uncertainties that may influence the
distributions after corrections and unfolding were quantified, and their
magnitudes are summarised in Table~\ref{tab:systematics}. The sources of
uncertainty and the methods used to estimate them were the following:
\begin{description}
\item[Trigger and vertexing:] The systematic uncertainties were found to be
  negligible.
\item[Track reconstruction:] The uncertainties in track reconstruction
  efficiency principally arise from imperfect knowledge of the 
  material in the inner detector. The new insertable B-layer and changes to
  pixel detector services in the $|\eta| > 1.5$ region add uncertainties not
  included in the inner-detector material assessments performed during Run~1:
  these have been evaluated by comparison of data to simulation, and by
  comparisons of simulations with different compositions of interacting
  particles. The result is an estimate of reconstruction efficiency uncertainty
  between 0.4\% and~1.5\%~\cite{Aad:2016mok}. An additional systematic uncertainty
  arises from possible biases and degradation in the \leadpt measurement:
  these effects were determined in Ref.~\cite{Aad:2016mok} and estimated to affect
  the tail of the \leadpt distribution by 4--5\%.
\item[Non-primary particles:] The systematic uncertainties were propagated
  by modification of track weights.  The systematic uncertainty on the selected
  non-primary track fraction was estimated to be 24\%, using variations of the fit range in
  the tail of the impact parameter distributions, and different MC
  generators and different shape assumptions for the extrapolation of the
  fraction from the side-bands to the signal region.
  The MC~variation is responsible for the most significant contribution to this
  uncertainty. The systematic uncertainty due to the fraction of charged strange
  baryons was derived using the deviations of generator predictions from
  \epos~\cite{Aad:2016mok}.
\item[Unfolding:] The systematic uncertainties associated with the HBOM
  unfolding have two distinct sources:\\[0.5ex]
  \textit{Non-closure:} The HBOM correction procedure is in principle
  independent of the Monte Carlo generator modelling. However, the method shows
  deficiencies in some regions and distributions. The relative size of the
  observed non-closure (i.e. non-reproduction of a known input) derived using
  \pythiaatwo was therefore applied as a correction. The size of the correction
  is included as a systematic uncertainty, everywhere less than 2.5\%. The
  measured transverse momentum distribution of the leading charged particle was
  found to be consistent with the result obtained using Bayes-inspired iterative
  unfolding~\cite{D'Agostini:1994zf} within the estimated bias.\\[1ex]
  \textit{Parameterisation:} The statistical uncertainty of the HBOM method was
  derived for each bin using Monte Carlo sampling: for each $\mathrm{HBOM}(k)$
  iteration, samples were drawn from a Gaussian distribution corresponding to
  the point uncertainty, and 1000 independent replica HBOM fits were
  performed. The resulting uncertainty contribution is derived from the
  replicas' central 68\% confidence interval. An additional systematic
  uncertainty, estimating the stability of the fit method, was derived by using
  different numbers of HBOM iterations $k$ and polynomial degrees $n$.
\end{description}
For the results presented in the following section, these independent sources of
systematic uncertainty have been combined in quadrature to form single
representative systematic uncertainty estimates. 

\begin{table}[tbp]
  \let\nr\numrange
   \centering
   \caption{Summary of systematic uncertainties for each class of UE observable, broken down by origin.}
   \label{tab:systematics}
   \vspace{1ex}
   \begin{tabular}{lllll}
      \toprule
       & \multicolumn{4}{c}{Range of values}\\
      \cmidrule(l){2-5}
      Observable                                            & Material & Non-primaries & Non-closure & Parameterisation \\
      \midrule
      \nch or \sumpt vs. \dphi                              & 0.9\%           & 0.6\%           & $0$--$0.6\%$ & $0$--$0.4\%$\\
      \nch or \sumpt vs. \leadpt                            & $0.5$--$1.0\%$  & $0.3$--$0.6\%$   & $ 0$--$2.5\%$ & $ 0$--$0.4\%$\\
      \meanmeanpt vs. \nch                                  & $0$--$0.5\%$    & $0$--$0.5\%$     & \multicolumn{2}{c}{---~~0.5\%~(combined)~~---}\\
      \meanmeanpt vs. \leadpt                               & $0$--$0.4\%$    & $0$--$0.3\%$     & \multicolumn{2}{c}{---~~0.5\%~(combined)~~---}\\
      \bottomrule
   \end{tabular}
\end{table}

\section{Results}
\label{sec:result}
\begin{figure*}[bp]
  \begin{center}
     \includegraphics[width=\picwtwores\textwidth]{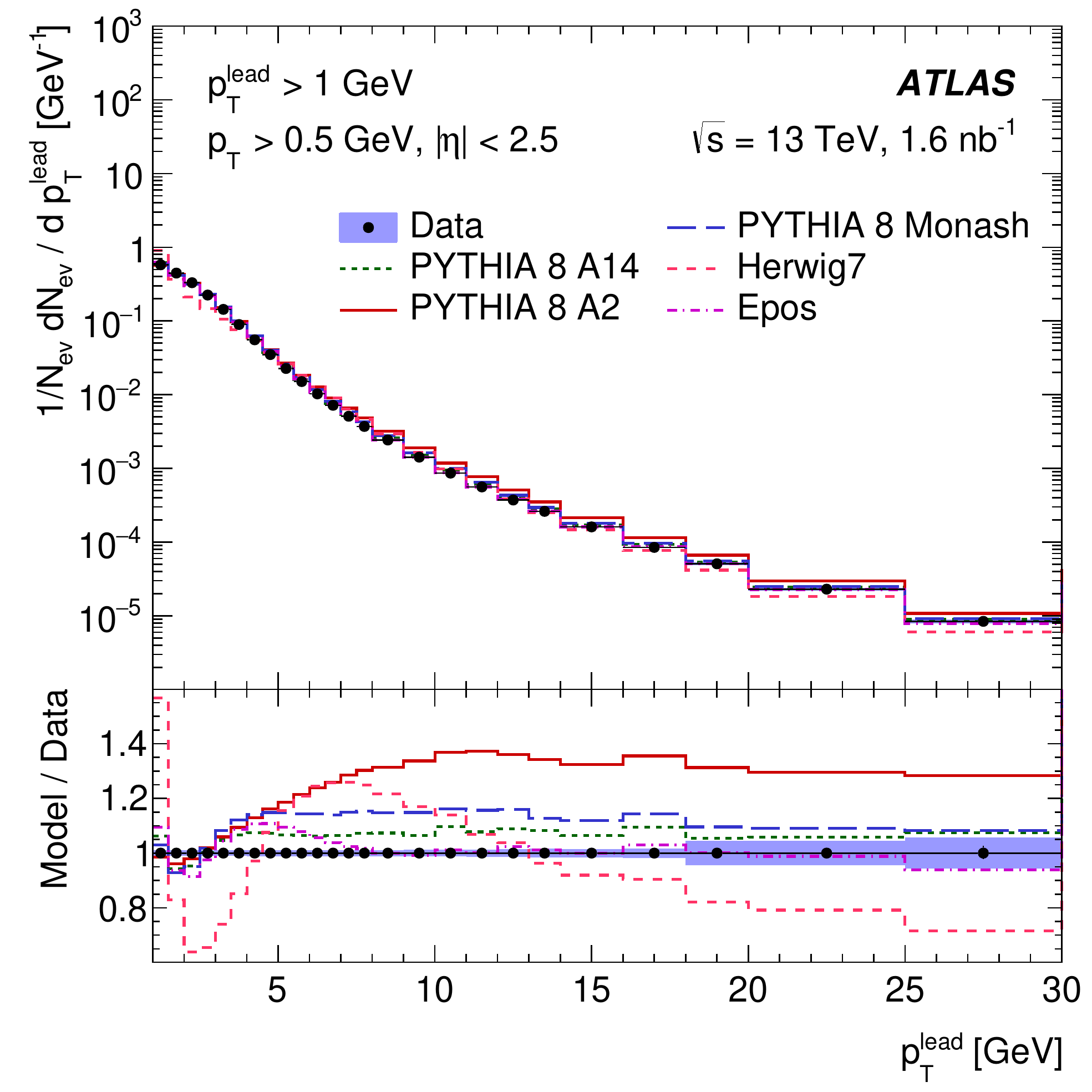}
  \end{center}
  \caption {Unit-normalised distribution of the transverse momentum of the
    leading charged particle, $\ptlead > \SI{1}{\GeV}$, compared to various
    generator models. The error bars on data points represent statistical
    uncertainty and the blue band the total combined statistical and systematic
    uncertainty.}
  \label{fig:leadpt}
\end{figure*}

\begin{figure*}[tp]
  \begin{center}
    \includegraphics[width=\picwtwores\textwidth]{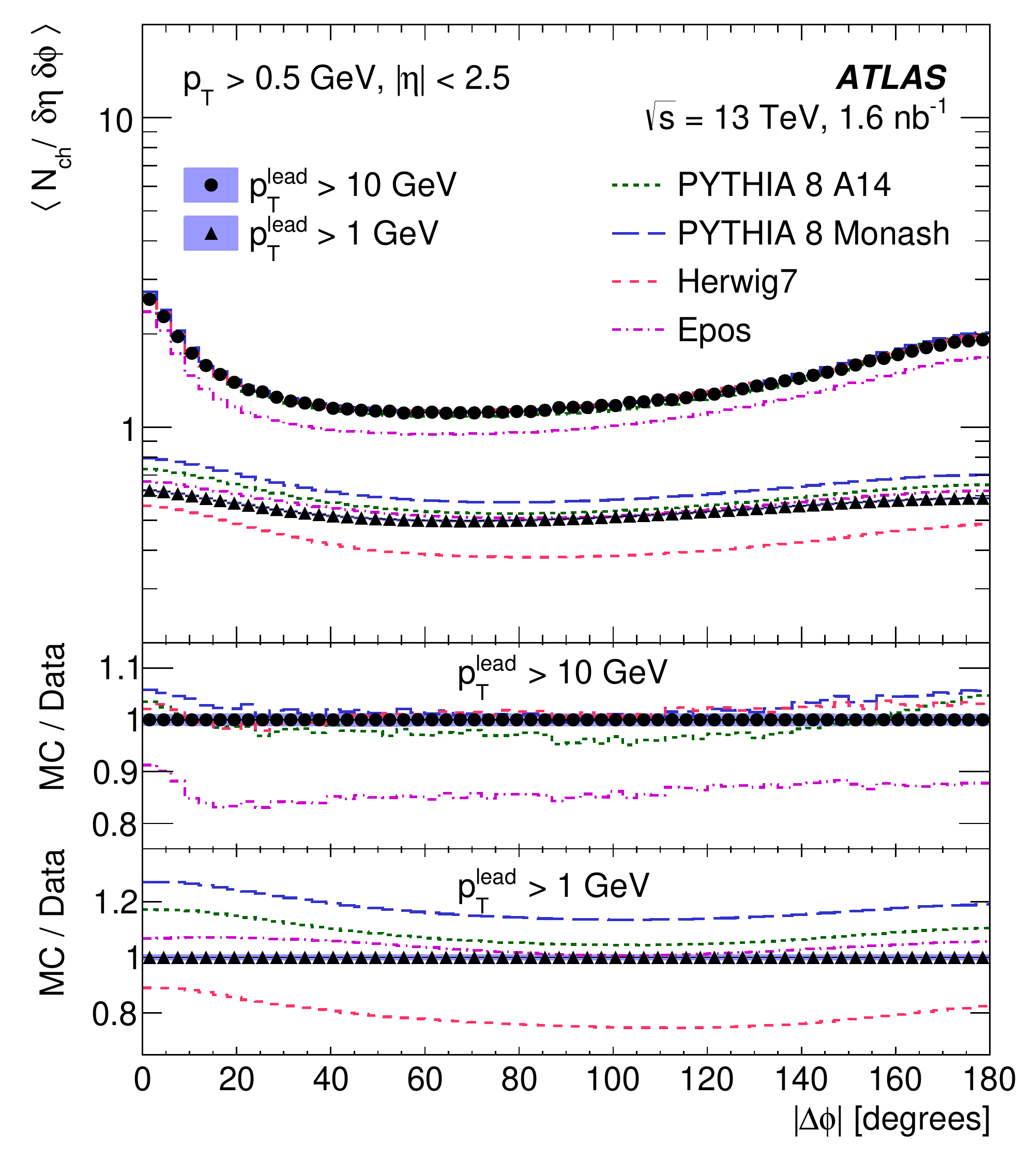}
    \includegraphics[width=\picwtwores\textwidth]{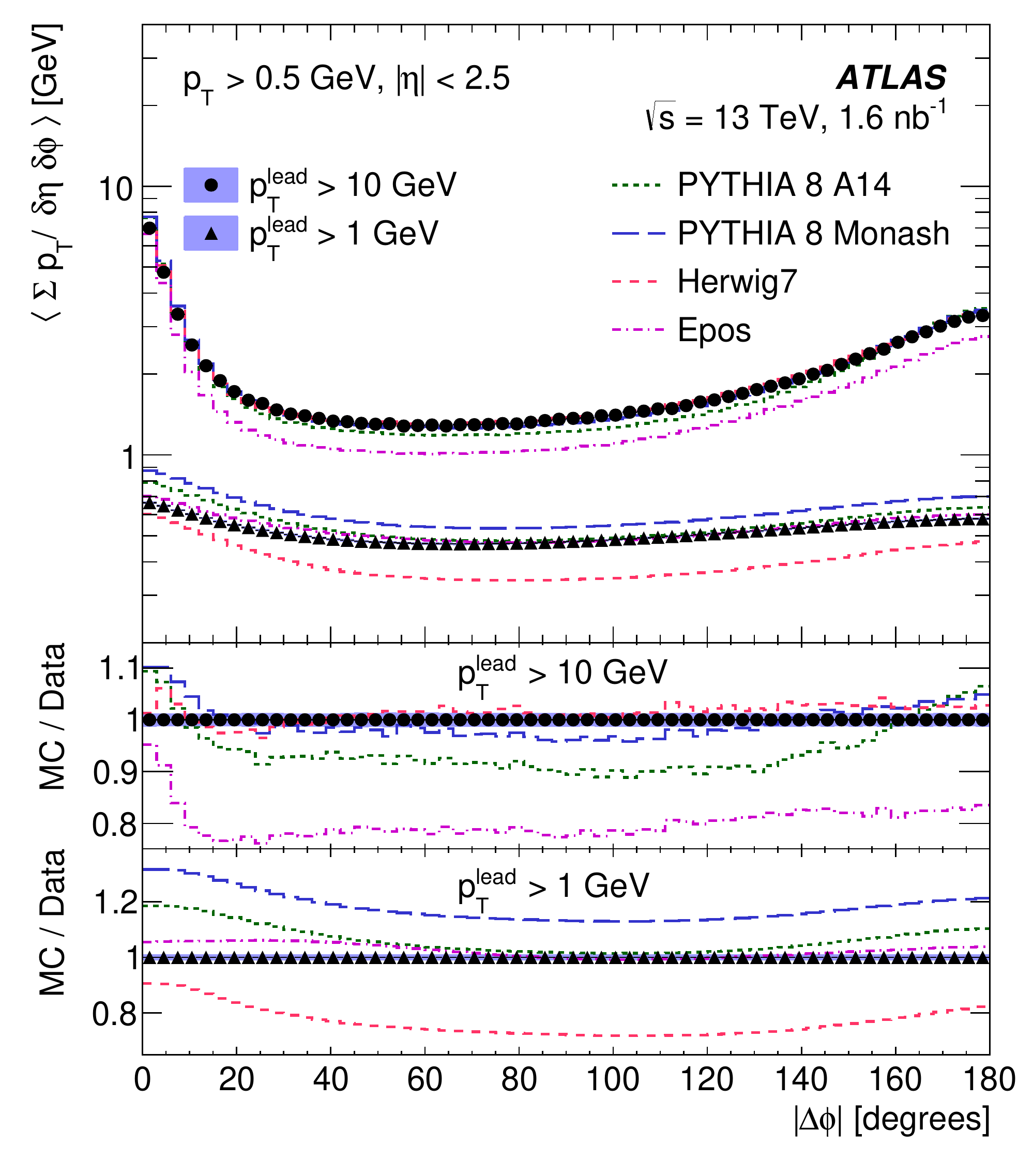}
  \end{center}
  \caption {Distributions of mean densities of charged-particle multiplicity \nch
    (left), and \sumpt (right) as a function of \Adphi (with respect
    to the leading charged particle) 
    for $\ptlead > \SI{1}{\GeV}$ and $\ptlead > \SI{10}{\GeV}$ separately, with
    comparisons to MC generator models.
    The error bars on data points represent statistical uncertainty and the blue
    band the total combined statistical and systematic uncertainty.}
  \label{fig:dphi}
\end{figure*}

\begin{figure*}[tp]
  \vspace*{-2ex}
  \begin{center}
    \includegraphics[width=\picwtwores\textwidth]{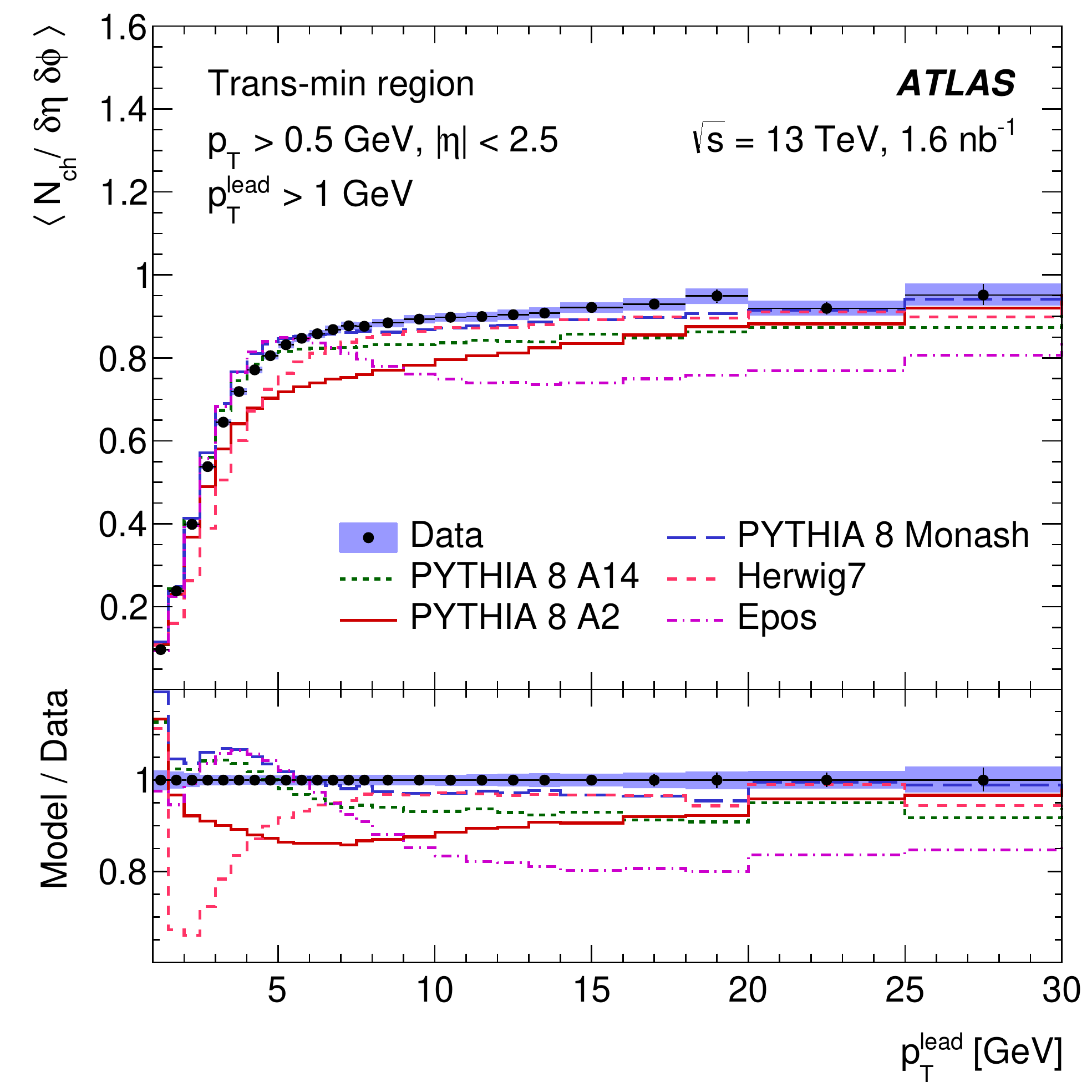}
    \includegraphics[width=\picwtwores\textwidth]{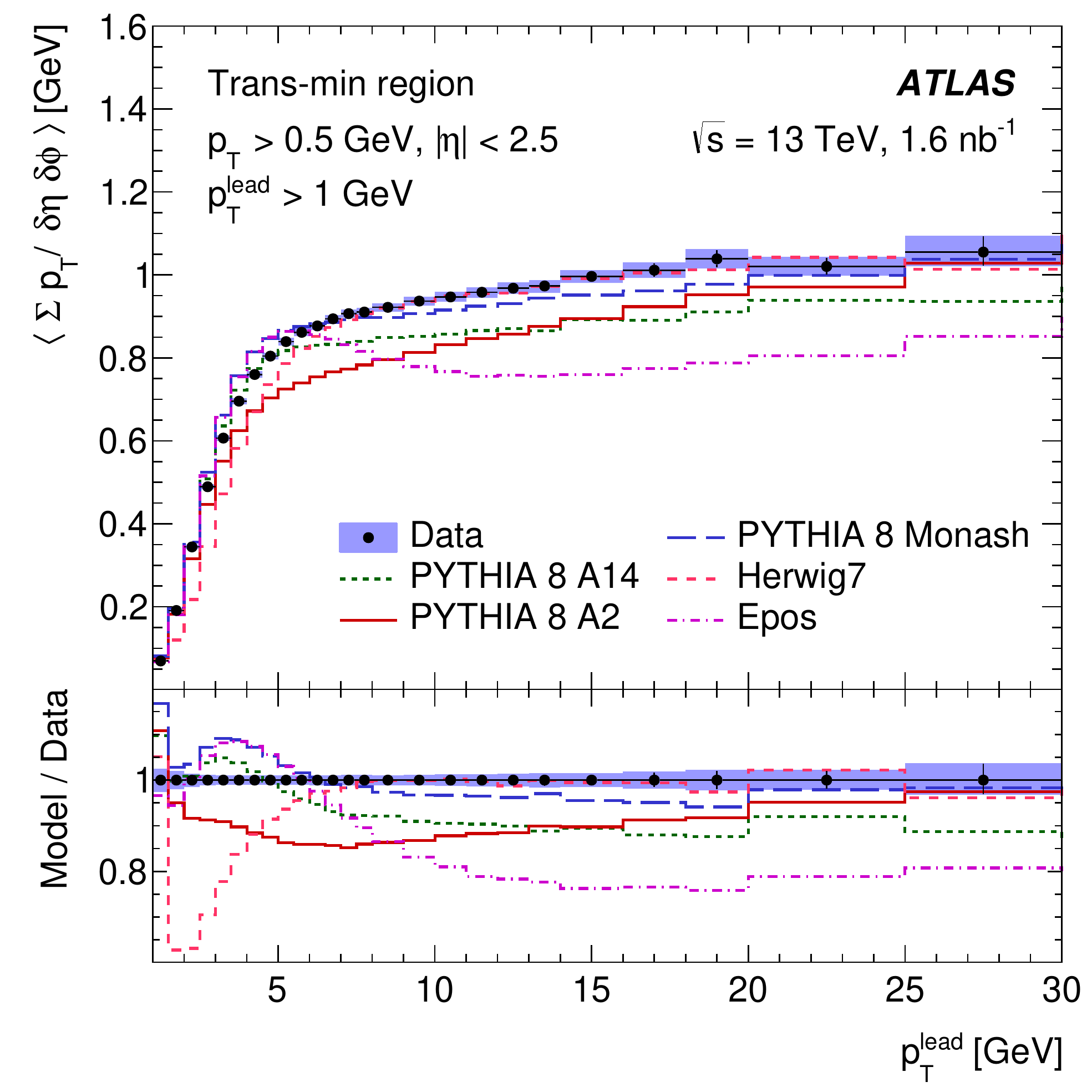}\\
    \includegraphics[width=\picwtwores\textwidth]{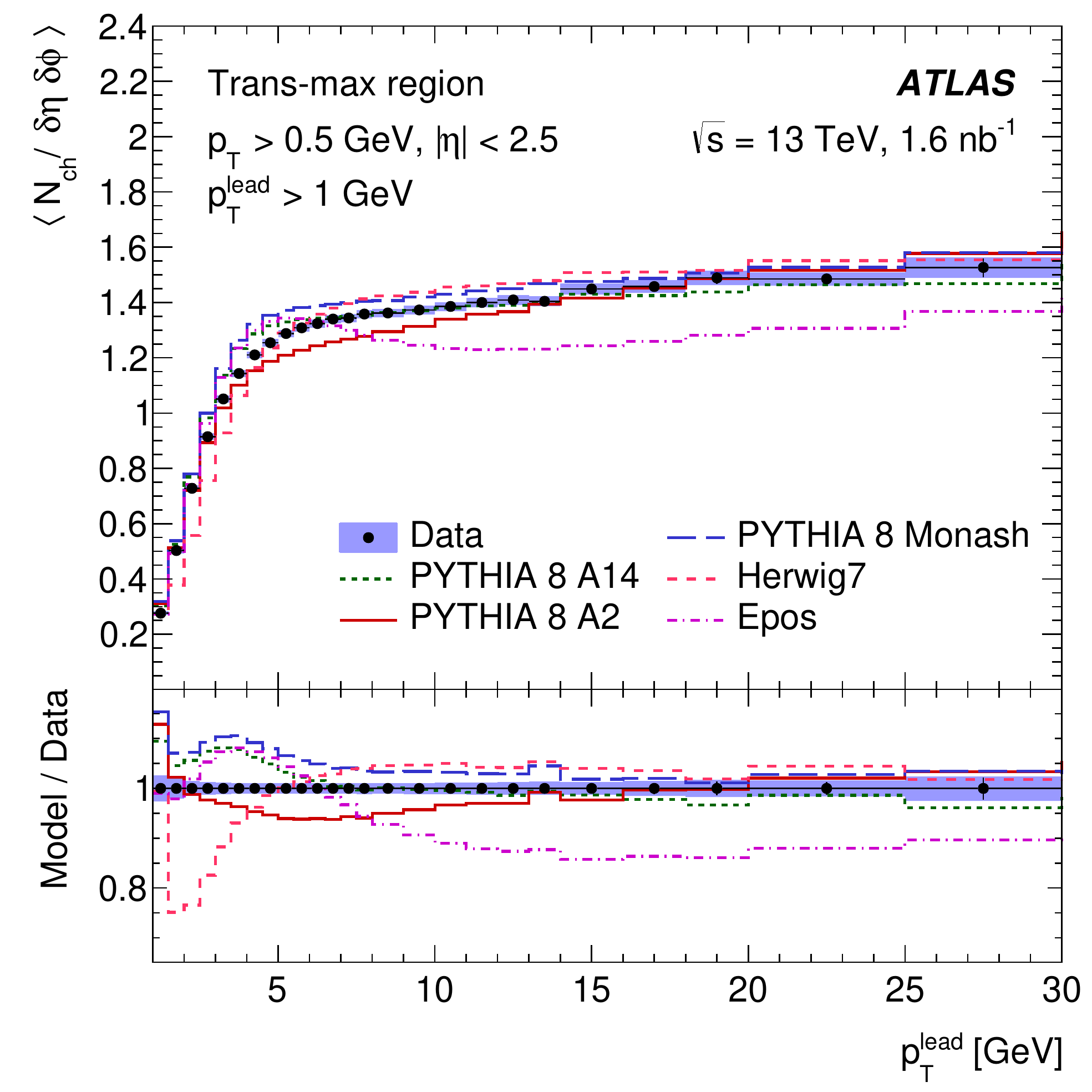}
    \includegraphics[width=\picwtwores\textwidth]{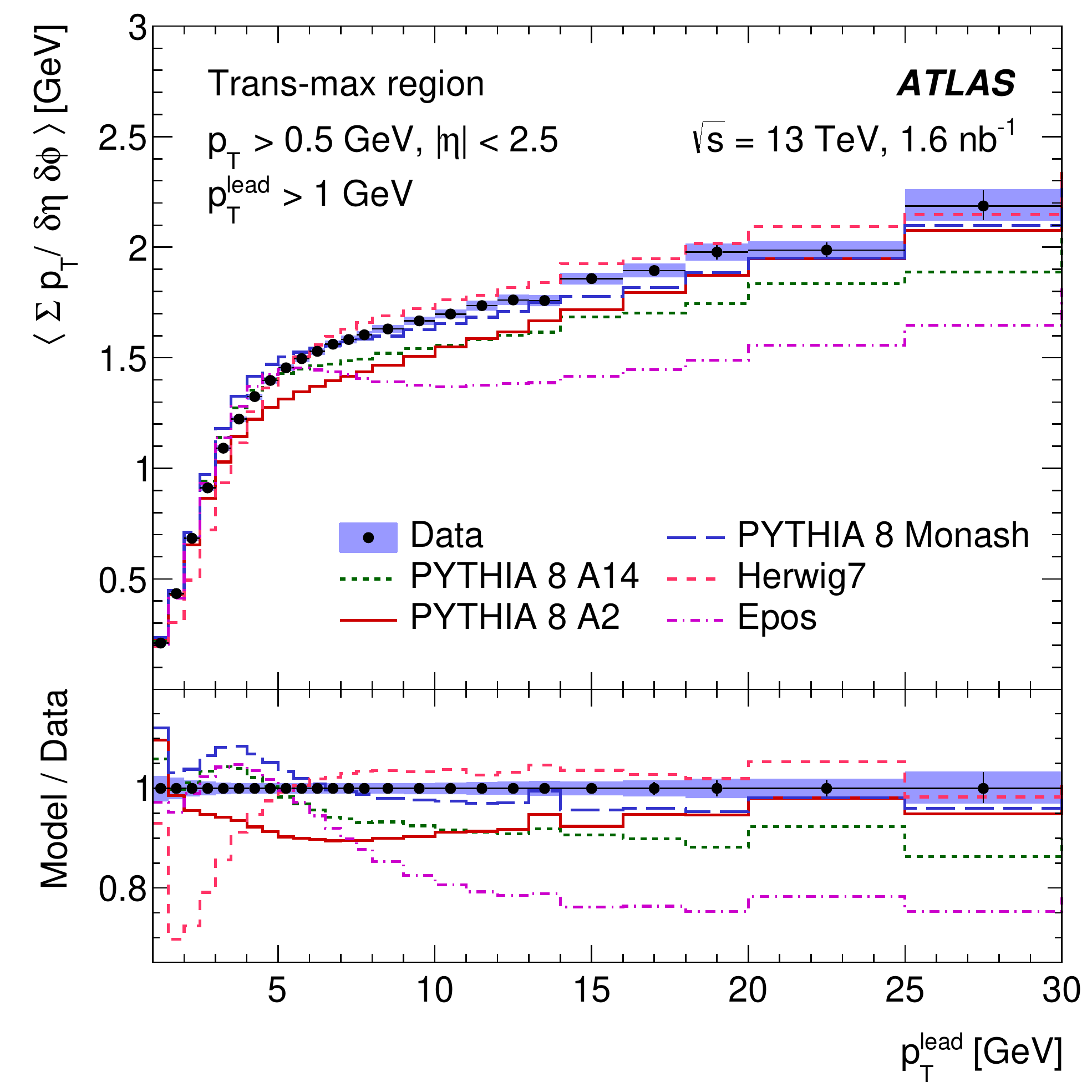}\\
    \includegraphics[width=\picwtwores\textwidth]{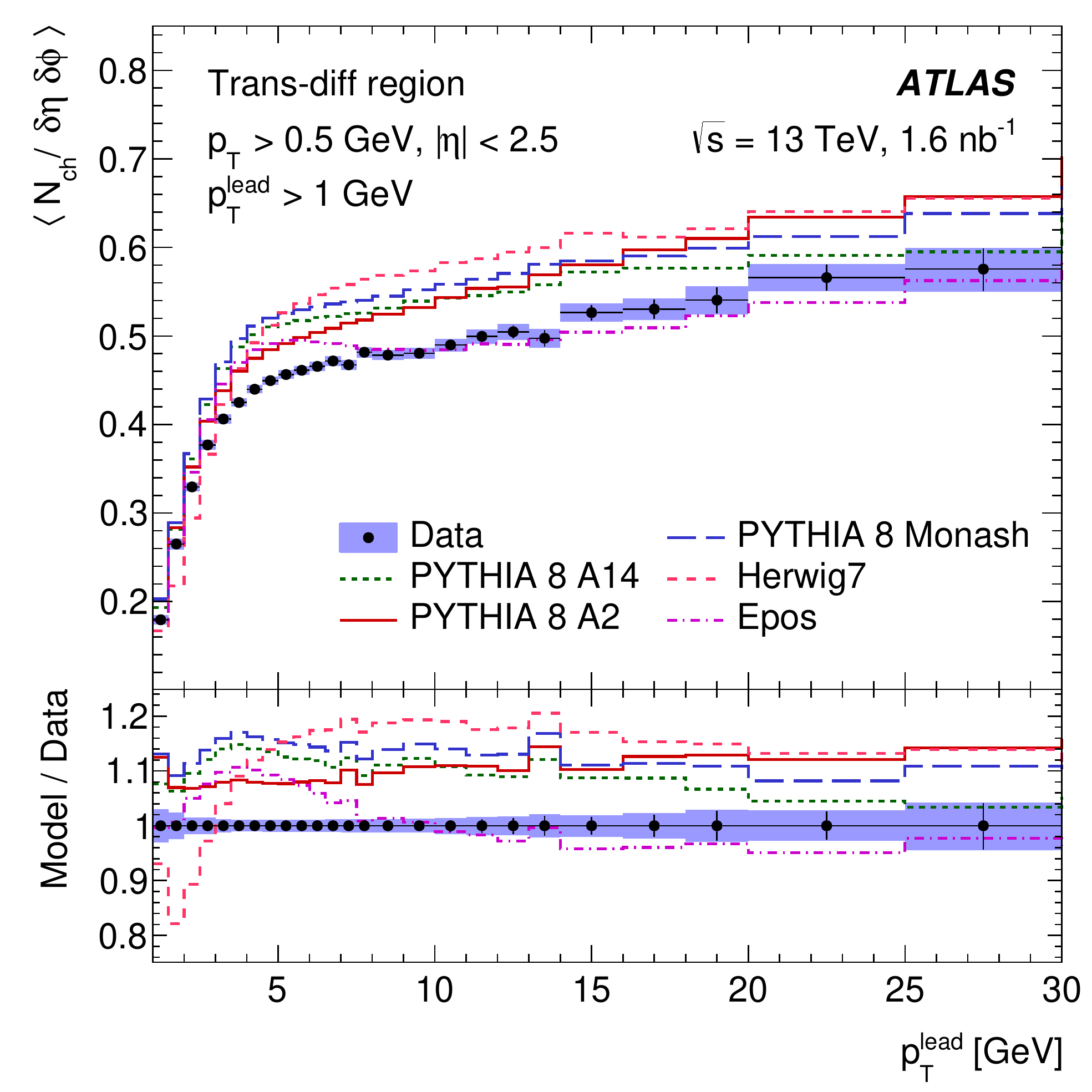}
    \includegraphics[width=\picwtwores\textwidth]{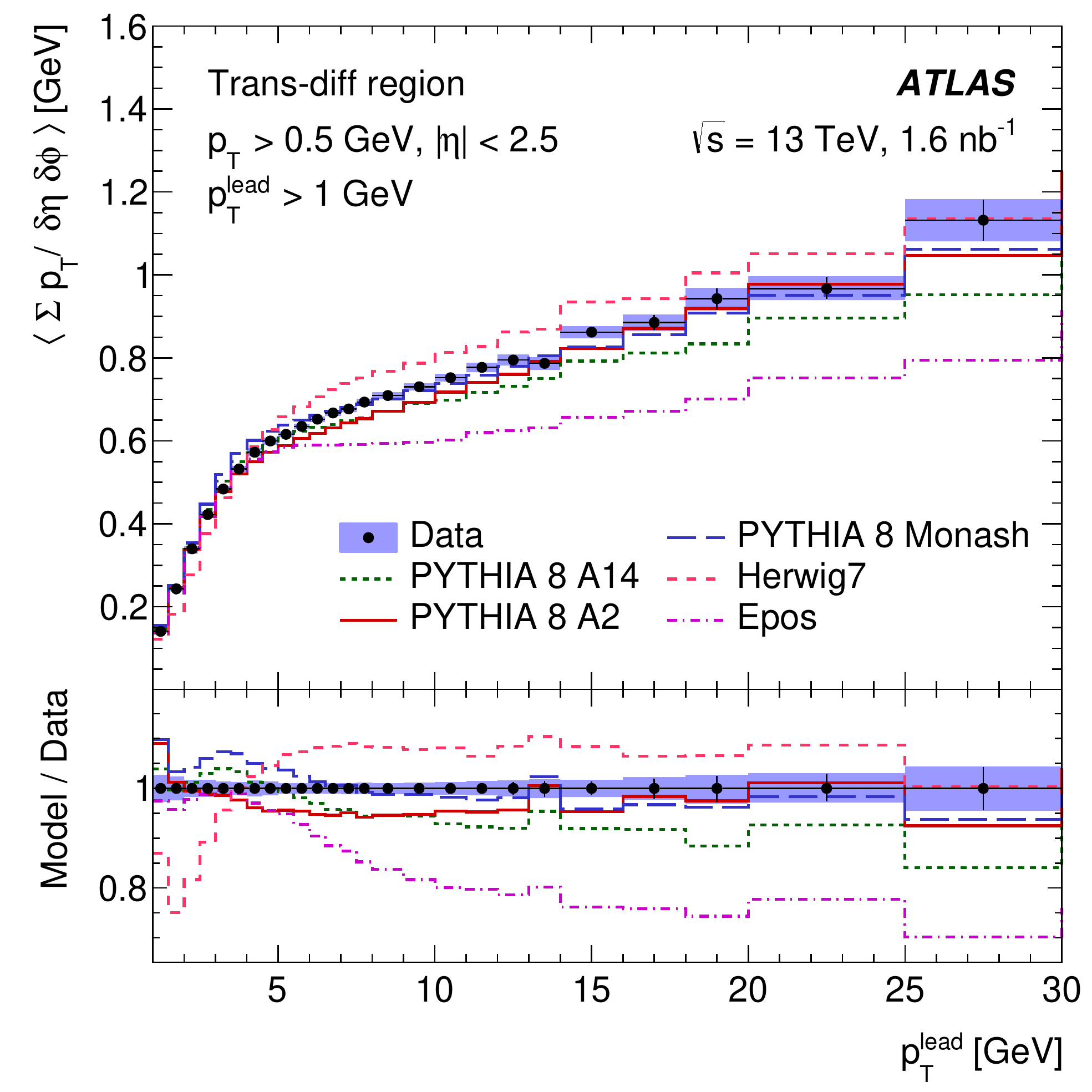}
  \end{center}
  \vspace*{-1.5em}
  \caption {Mean densities of charged-particle multiplicity \nch (left) and \ptsum
    (right) as a function of leading charged-particle \pt, in the trans-min
    (top), trans-max (middle) and trans-diff (bottom) azimuthal regions.
    The error bars on data points represent statistical uncertainty and the blue
    band the total combined statistical and systematic uncertainty.}
  \label{fig:densities_transmmd}
\end{figure*}

\begin{figure*}[tp]
  \begin{center}
    \includegraphics[width=\picwtwores\textwidth]{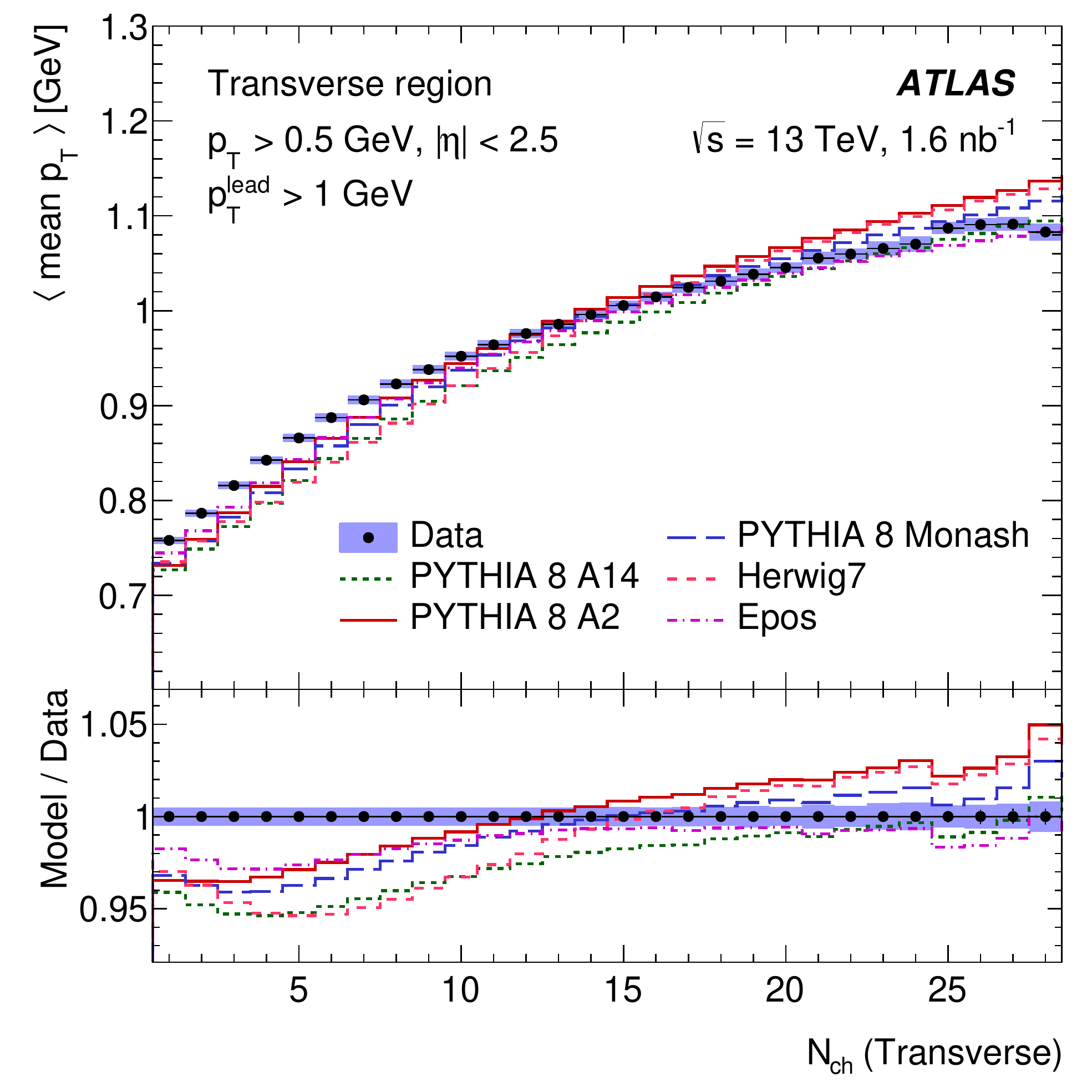}
    \includegraphics[width=\picwtwores\textwidth]{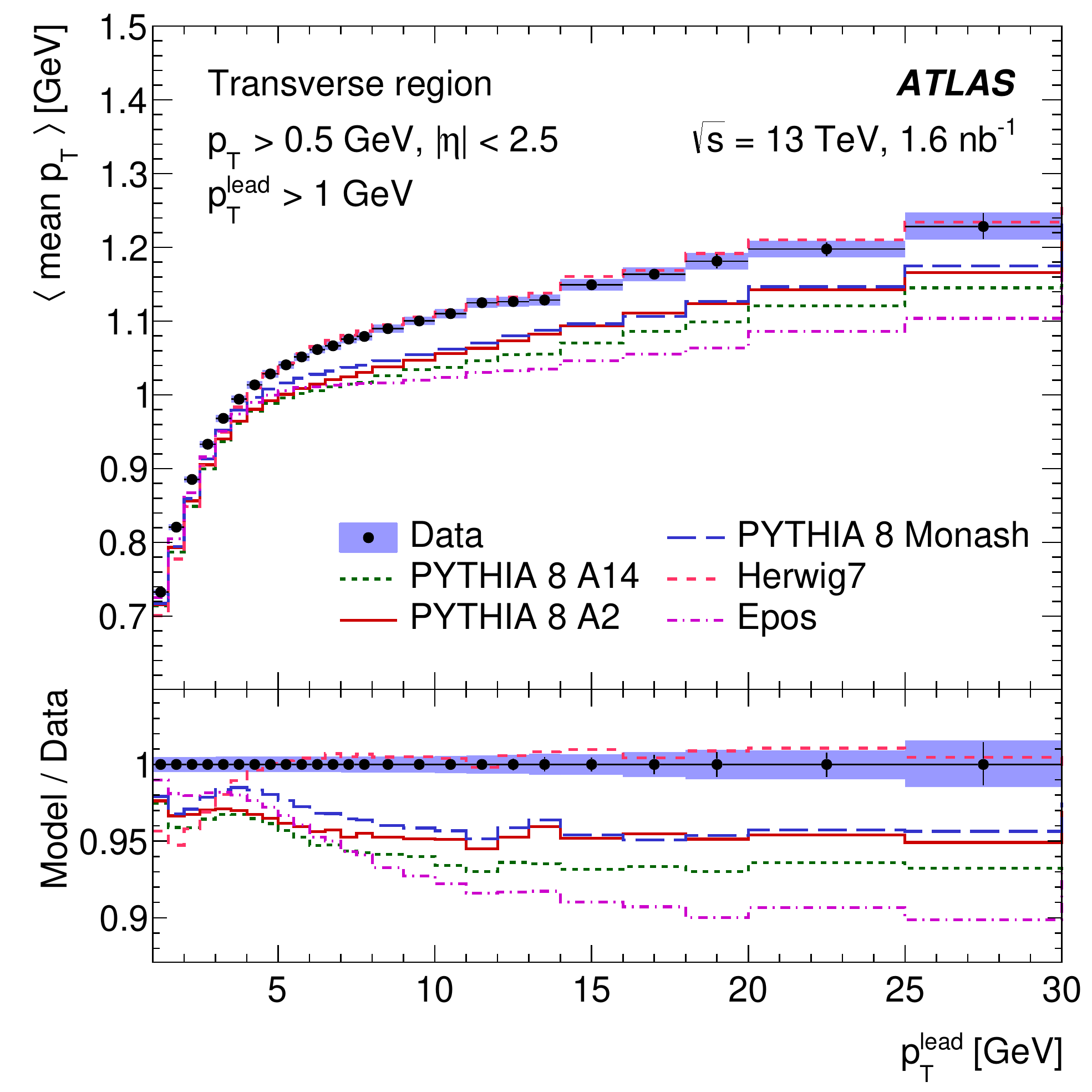}\\
    \includegraphics[width=\picwtwores\textwidth]{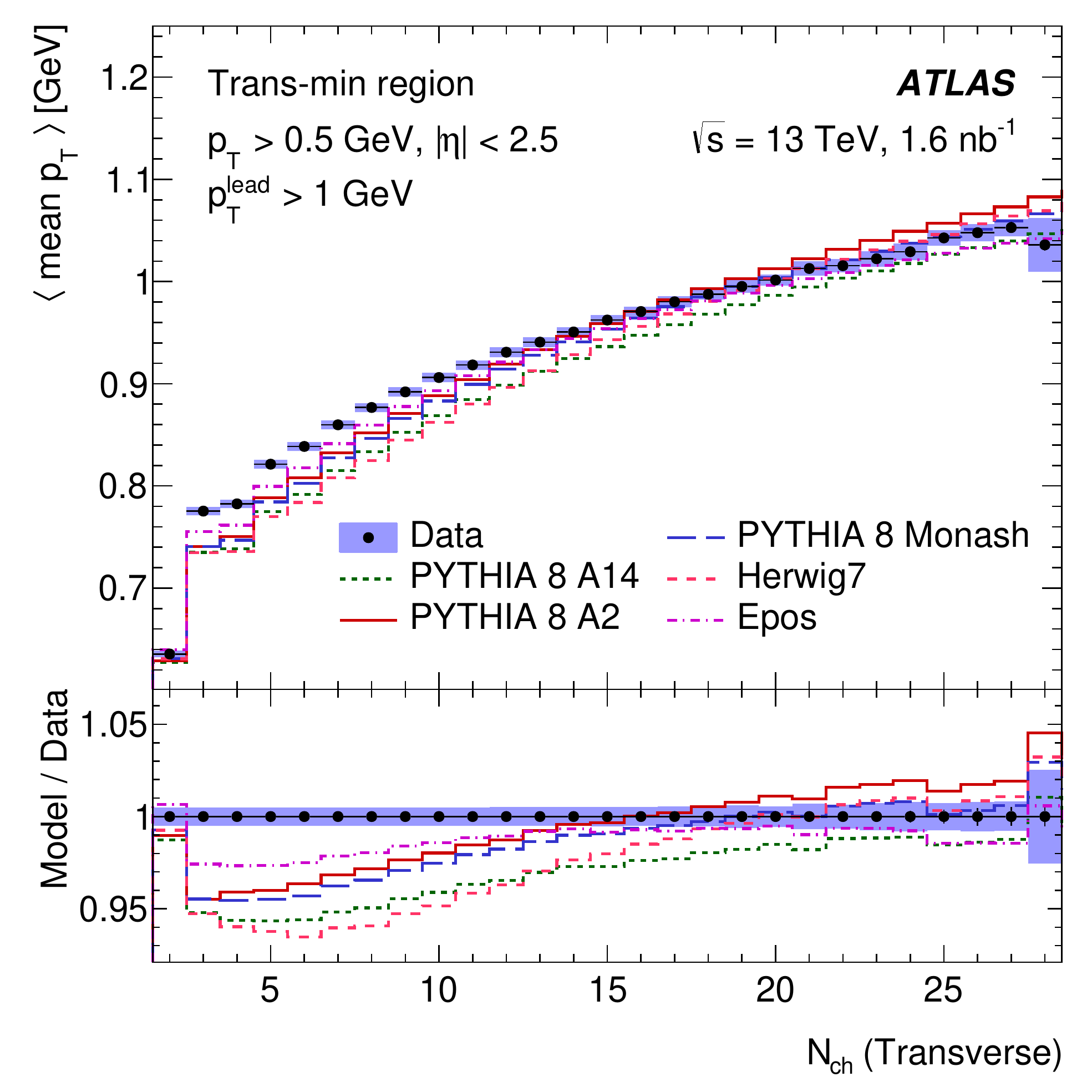}
    \includegraphics[width=\picwtwores\textwidth]{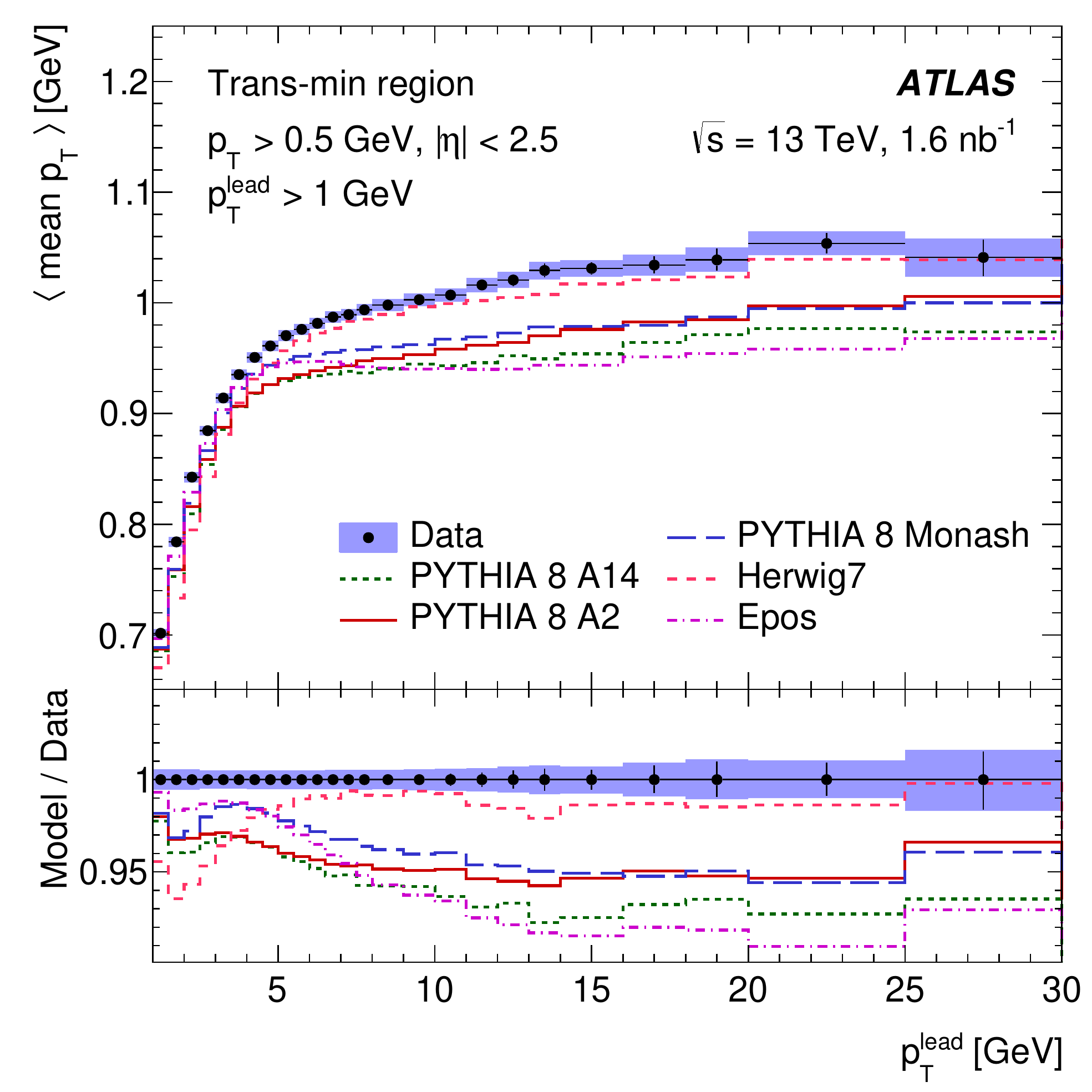}\\
    \includegraphics[width=\picwtwores\textwidth]{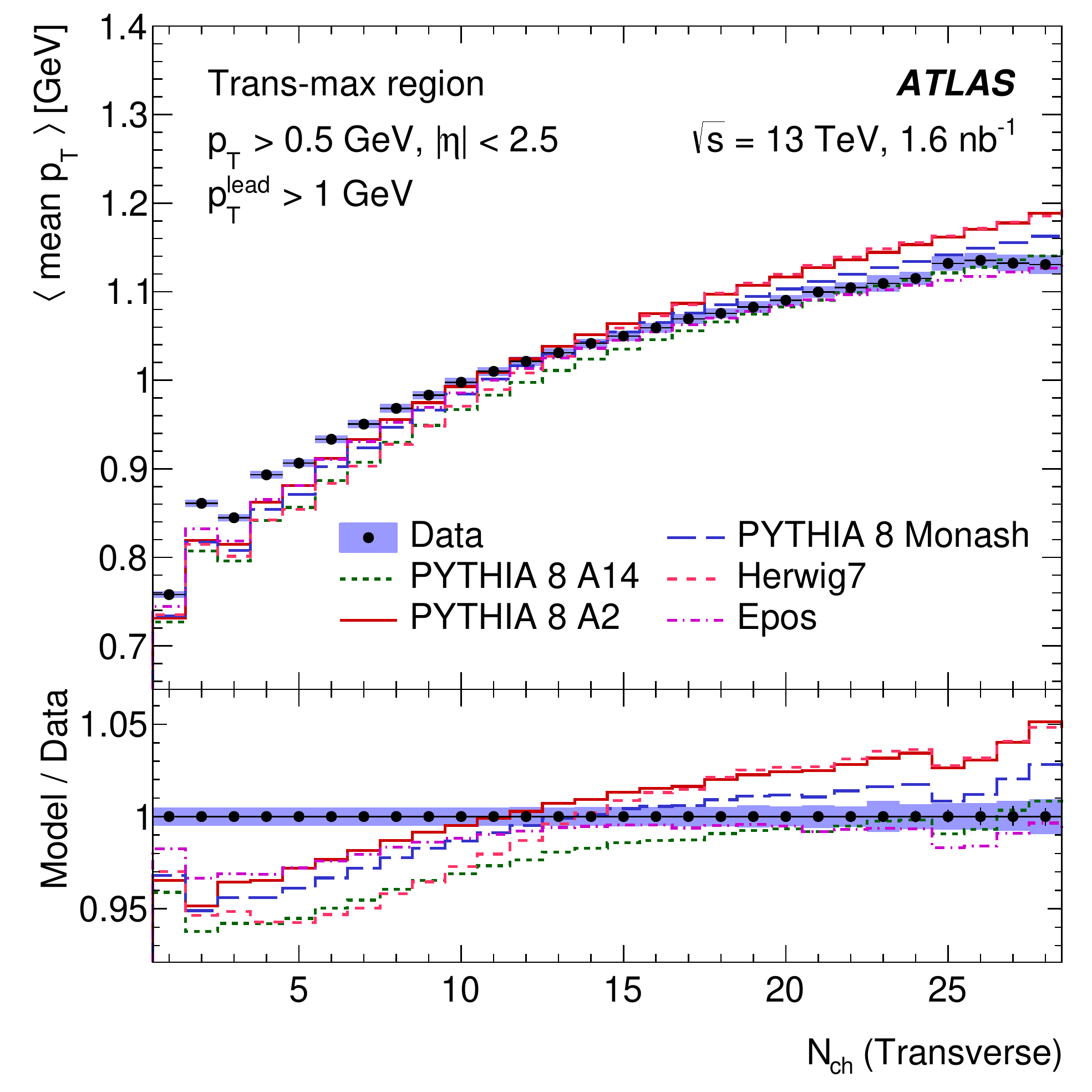}
    \includegraphics[width=\picwtwores\textwidth]{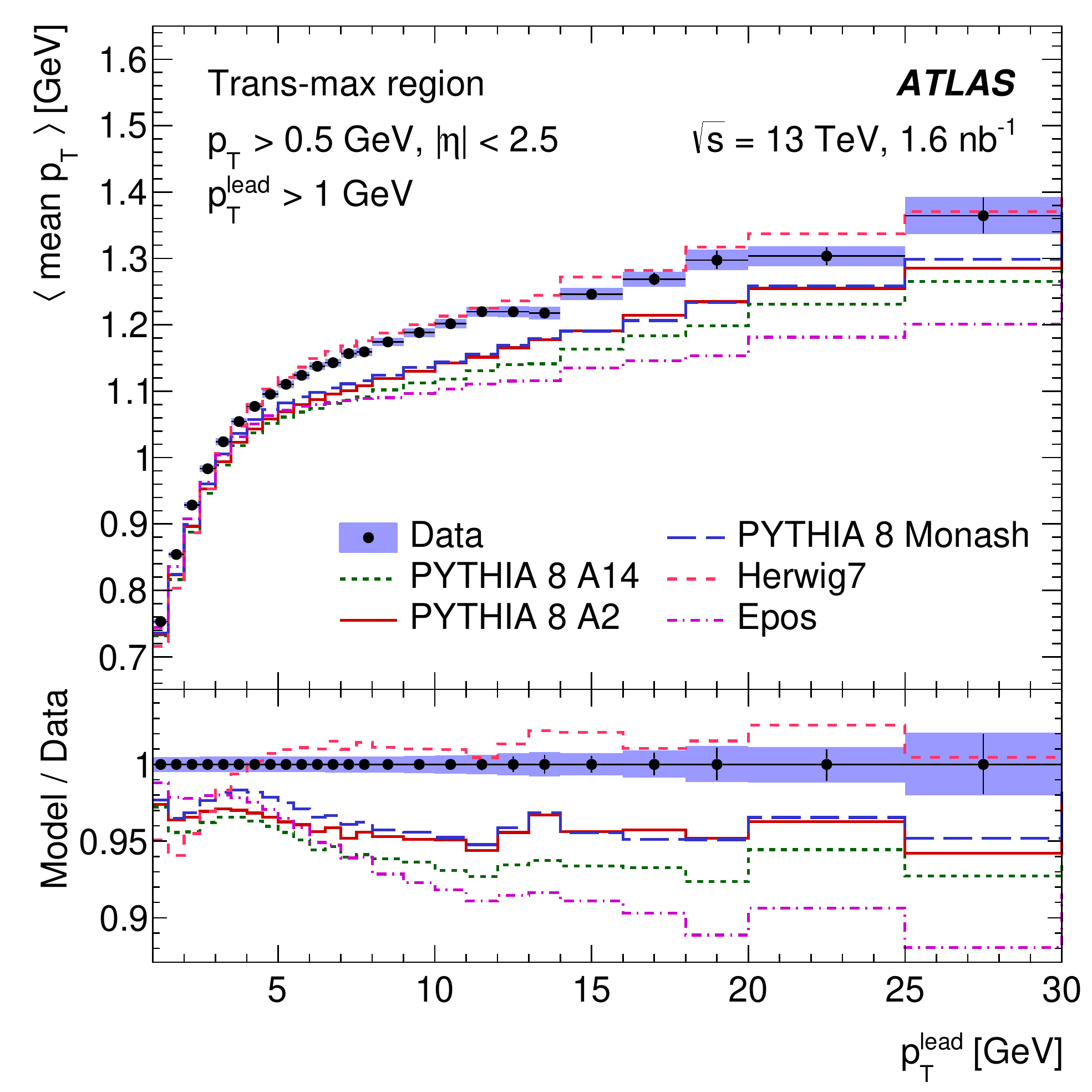}
  \end{center}
  \vspace*{-1em}
  \caption{Mean charged-particle average transverse momentum as a function of
    charged-particle multiplicity in transverse region
    $\nch\mathrm{(Transverse)}$~(left) and as a function of
    \leadpt~(right), for each of the transverse~(top), trans-min~(middle) and
    trans-max~(bottom) azimuthal regions. The error bars on data points represent
    statistical uncertainty and the blue band the total combined statistical and systematic uncertainty.}
  \label{fig:meanpt_tmm}
\end{figure*}

The unfolded distributions and their main features are discussed here, and in
Figures~\ref{fig:leadpt}, \ref{fig:dphi}, \ref{fig:densities_transmmd}
and~\ref{fig:meanpt_tmm} are compared to model predictions
from \pythiaeight (the A14, A2 and Monash tunes), \herwigseven, and \epos. These
model predictions were obtained using the analysis' associated Rivet
routine~\cite{Buckley:2010ar}.



\paragraph{Leading charged-particle \pt}

Figure~\ref{fig:leadpt} shows the unit-normalised
distribution of events
with respect to the transverse momentum of the leading charged particle,
\ptlead. This is a steeply falling distribution, with a change of slope
for $\ptlead \gtrsim \SI{5}{\GeV}$: a form which is broadly modelled by all generators. The
\pythiaeight~A14 and Monash tunes, as well as \epos, model the distribution within 15\% out to
$\ptlead = \SI{30}{\GeV}$, while the \pythiaeight A2 minimum-bias 
tune predicts too hard a spectrum in the high $\ptlead$ tail.  \herwigseven
peaks strongly at the lowest \ptlead and alternates between under- and
over-shooting the data at higher scales, finally producing a softer tail
than seen in data.

\paragraph{Angular distributions versus leading charged particle}

Figure~\ref{fig:dphi} shows the mean multiplicity and \sumpt distributions as a
function of azimuthal angle with respect to the leading particle for different
\ptlead requirements. Two event selections are shown here: the
$\ptlead > \SI{1}{\GeV}$ cut common to all 
observables, and a harder $\ptlead > \SI{10}{\GeV}$ requirement.

The difference between these two selections illustrates the transition from relatively
isotropic minimum-bias scattering to the emergence of hard partonic
scattering structure and hence a dominant axis of energy flow. 
This event structure with least activity perpendicular to the leading-object
axis, i.e. away from $\dphi = $~\SIlist{0;180}{\degree}, is seen for both
selections and both observables but is much stronger for the event subset with
the higher $\ptlead > \SI{10}{\GeV}$ requirement: this demonstrates the evolution of
event shape as a hard scattering component develops.

There is no clear ``best'' MC model for these observables. \epos performs best in
the more inclusive $\ptlead > \SI{1}{\GeV}$ selection, followed by
\pythiaeight~A14; \herwigseven significantly undershoots while \pythiaeight
Monash is everywhere above the data. But in the hard-scattering
$\ptlead > \SI{10}{\GeV}$ event selection, \herwigseven and Monash perform best,
with a slight undershoot from \pythiaeight~A14 and a large one from \epos. These orderings
apply to both the \nch and \ptsum variables, although to different extents.

\paragraph{\nch and \ptsum densities in transverse/towards/away regions}

\begin{figure*}[tp]
  \begin{center}
    \includegraphics[width=\picwtwores\textwidth]{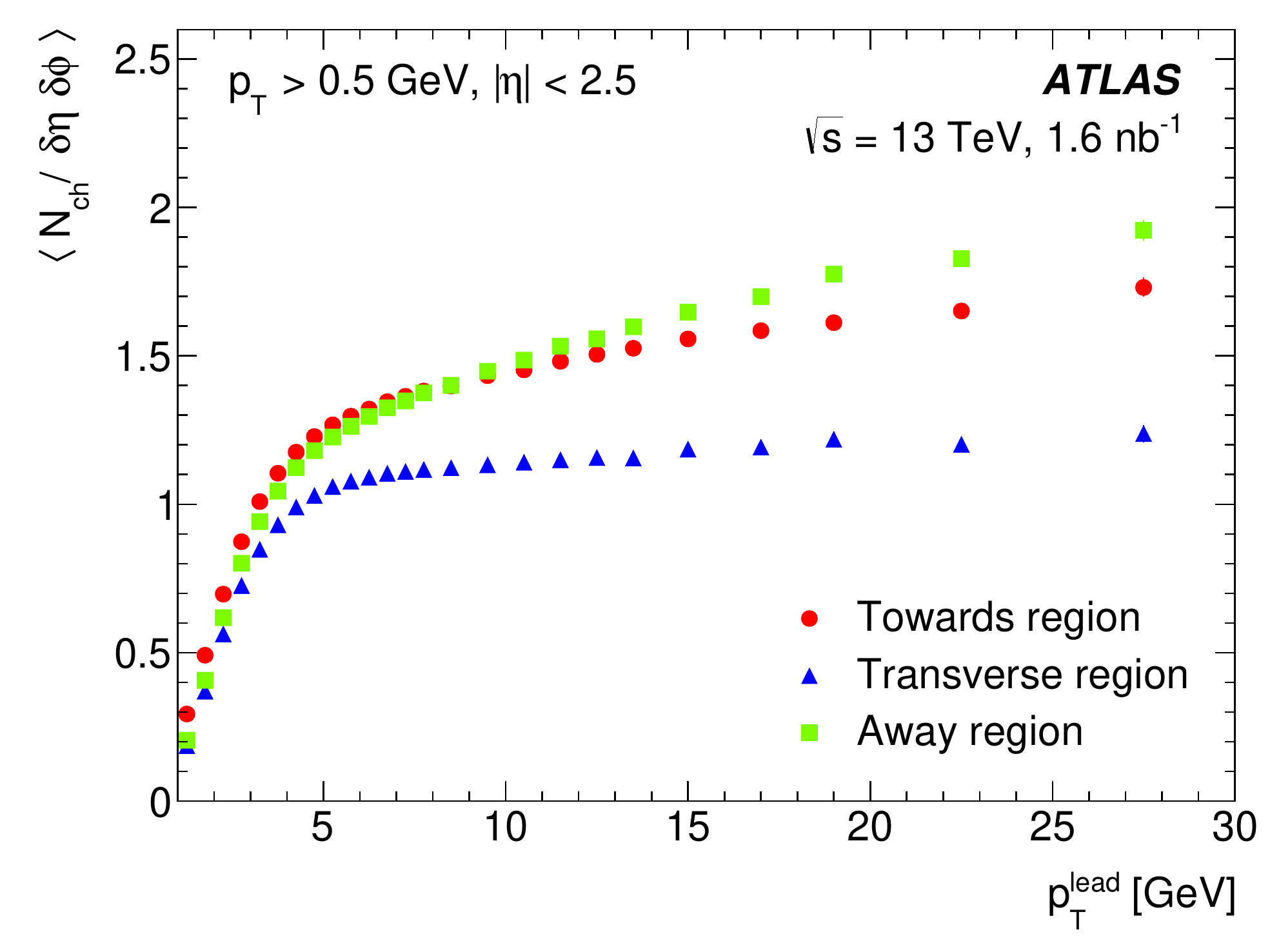}
    \includegraphics[width=\picwtwores\textwidth]{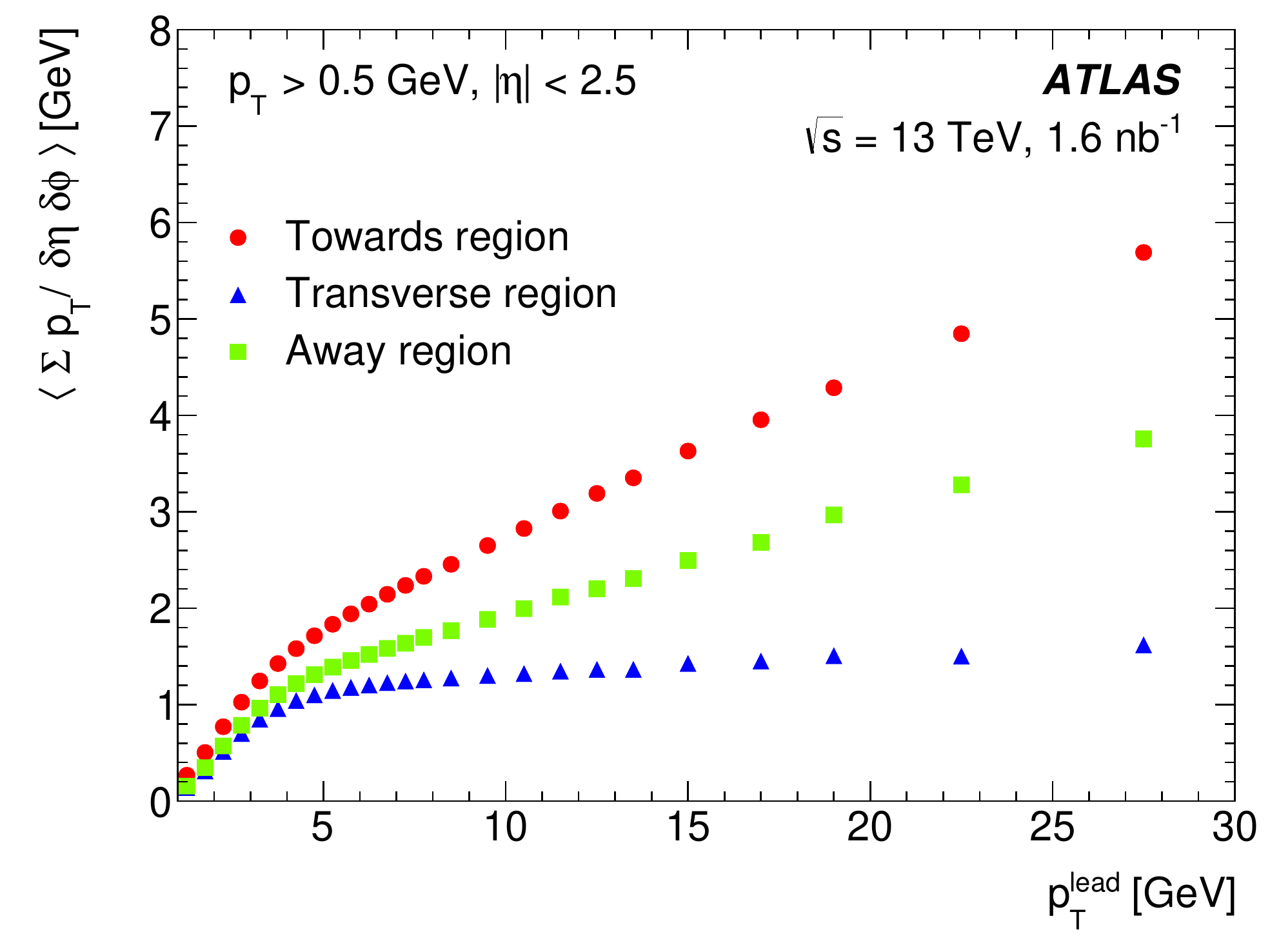}
  \end{center}
  \caption {Mean \etaphi densities of charged-particle multiplicities (left) and
    \ptsum (right) as a function of the transverse momentum of the leading
    charged particle in the transverse, towards, and away azimuthal regions.
    The error bars, which are mostly hidden by the data markers,
    represent combined statistical and systematic uncertainty.}
  \label{fig:densities_trans}
\end{figure*}

Figure~\ref{fig:densities_trans} shows the evolution of the mean 
charged-particle multiplicity and \ptsum
densities with the \pt of the leading charged particle, from 1 to
\SI{30}{\GeV}. For both observables, the towards, transverse, and away regions
are shown overlaid for ease of comparison.

The primary feature is the general shape seen in all curves: starting close to
zero at low \ptlead, there is first a very rapid rise in
activity in which the three regions are not strongly distinguished, then an
abrupt transition at $\ptlead \approx \SI{5}{\GeV}$ above which the three
regions have quite distinct behaviours. The initial rapid rise to a roughly
stable value of \num{\sim 1} charged particle or \SI{\sim 1}{\GeV} per unit
\etaphi area is known as the ``pedestal effect''. In modelling terms this reflects a
reduction of the $pp$ impact parameter with increasing \ptlead, and hence the
transition between the minimum-bias and hard-scattering regimes.

Secondly, the shape of the transverse region is different from the other two
in both observables -- it almost completely plateaus after
$\ptlead \approx \SI{5}{\GeV}$, while the towards and away regions continue to
rise nearly linearly with \ptlead. This characteristic feature of UE profile
observables is the empirical demonstration of the azimuthal-region paradigm for
UE analysis: the hard process dominates the towards and away regions, which
continue to increase in activity as the hard-process scale grows, but the
transverse region is relatively unaffected. This is consistent with the pedestal effect 
where the overlap between colliding protons is complete and hence any further
growth is due to connections to, or contaminations from, the hard process rather
than more MPI scattering.

An interesting feature is that for $\ptlead \gtrsim \SI{7}{\GeV}$ the \emph{away}
region actually becomes the one with highest charged-particle multiplicity,
despite not containing the highest-\pt charged particle. By the \ptsum density
measure, however, the towards region is unambiguously the most active region for
all \ptlead values.

\paragraph{\nch and \ptsum densities in trans-min/max/diff regions}

Figure~\ref{fig:densities_transmmd} focuses on the UE-dominated transverse
region, and its per-event specialisations trans-min, -max, and
-diff. Between these, the trans-min is the most sensitive to MPI effects,
i.e.~the pedestal, while the trans-max includes both MPI and hard-process
contaminations: the trans-diff is hence the clearest measure of those
contaminations. The comparisons to MC models are again made in these plots.

There is significant variation in performance between the models, with
\pythiaeight's Monash tune and \herwigseven giving the best description of data in the
plateau region of trans-min, followed within 10\% by the other \pythiaeight
tunes. \epos, however, slightly overestimates in the ``ramp'' region
to the pedestal effect plateau, and on the plateau it underestimates
the pedestal height by around 20\%. \herwigseven and \pythiaeight~A2 both
mismodel the transition, with a severe undershoot for \herwig below \ptlead of
\SI{5}{\GeV}, and a milder but broader undershoot from A2 which extends up to
$\ptlead \approx \SI{20}{\GeV}$.

The predictions cluster together more tightly in the process-inclusive trans-max region,
with all generators other than \epos providing a description of the \nch density
data within a few percent for $\ptlead \geq \SI{10}{\GeV}$; \epos continues to
undershoot the data significantly. \pythiaeight~A14 also significantly undershoots the \ptsum
density data, by around 10\% as compared to \epos's 20\%. Looking in trans-diff
gives a clearer view of how non-MPI contributions to the UE are modelled, with
mostly flat \num{\sim 10}\% overshoots from all models other than \epos in
\nch density, and a spread of performance in describing the \ptsum density
evolution. In the latter the best performance comes from the \pythiaeight Monash and
A2 tunes, with \herwigseven and \pythiaeight~A14 both wrong by 5--10\% but in opposite directions
-- \epos's prediction is again separated from the ATLAS data by more than 20\%.

There is no obvious best model for all observables, but the \pythiaeight
Monash tune agrees well with the data in all observables other than trans-diff
\nch density, and \herwigseven has comparable performance for hard-scattering
events away from the ``minimum-bias region'' of $\ptlead < \SI{5}{\GeV}$. The
\pythiaeight~A14 tune, used for much of the hard-process simulation in ATLAS,
predicts activity 5\%~to~10\% 
below the data, indicating that some
re-tuning for \SI{13}{\TeV} event modelling may yield performance
benefits. \epos is not able to model the level of underlying-event activity well
in events with higher \ptlead.

\paragraph{Mean transverse momentum in transverse \& trans/min/max regions}

The per-event mean transverse momentum of charged particles in the transverse
azimuthal regions is of interest since it illustrates the balance in UE physics
between the \sumpt and multiplicity observables. This balance is affected in
some MC models by colour-reconnection or -disruption mechanisms, which
stochastically reconfigure the colour structures in the hadronising system into
energetically favourable states and typically increase the \pt per particle. Measurements of
the correlations of mean~\pt with other event features are an important input to
constrain such \textit{ad hoc} models. Figure~\ref{fig:meanpt_tmm} shows, for
the transverse and trans-min/max azimuthal regions, the correlations of the
average per-event mean transverse momentum, \meanmeanpt, with the transverse region's
charged-particle multiplicity and the event's \ptlead.

\afterpage{\clearpage}

The correlation with transverse charged-particle multiplicity \nch{} is a
correlation between two different ``soft'' properties. This distribution is
modelled to within 5\% for all \nch by most of the generators and in all
transverse region variants, but in all cases an underestimation of \meanmeanpt
is visible until $\nch \gtrsim 15$. The best modelling is from \epos, whose
maximum undershoot is \num{\sim 3}\% at low \nch but which follows the data
closely in all region definitions for higher transverse multiplicities. The
Monash tune is the best-performing \pythiaeight configuration, with a
performance similar to that of \epos, while \pythiaeight~A14 undershoots in the
low-multiplicity events and \pythiaeight~A2 overshoots at high-multiplicities --
notably the regions not included in each tune's construction. \herwigseven shows
the largest variations, from a \num{\sim 7}\% undershoot at low \nch to a 5\%
overshoot at $\nch \approx 30$.

However, \herwigseven performs better than all the other generators when considering the
correlation between transverse region \meanmeanpt and 
\ptlead.  As seen in Figure~\ref{fig:densities_transmmd}, the transverse \sumpt
does not reach as flat a plateau as does \nch with increasing \ptlead, and hence
the event-wise \meanmeanpt increases with \ptlead. \herwig's behaviour in
this observable is within 1\% of the data except in the ``minimum bias''
$\ptlead \lesssim \SI{5}{\GeV}$ phase space (where the \herwig model is not
expected to work), while all \pythiaeight tunes undershoot the data by between
5\%~and~10\%, and \epos even more so.

\FloatBarrier

\section{Conclusions}
\label{sec:conclusion}
We have presented several distributions sensitive to properties of the
underlying event, measured by the ATLAS experiment at the LHC using
\SI{1.6}{\per\nano\barn} of low-luminosity \SI{13}{\TeV} $pp$ collision events.
The measured observables are defined using charged
tracks and have been corrected to the level of primary charged particles.
Correction of the results for detector effects was performed via a
combination of tracking and vertexing efficiency weighting and HBOM unfolding,
with the latter accounting for re-orientation and similar indirect effects not
accounted for by the reweighting. Systematic uncertainties are included
for the effects of all data processing steps, including the unfolding.

The analysis observables follow the established strategy in which each event is azimuthally
segmented into ``towards'', ``transverse'', and ``away'' regions with respect to
the highest-\pt (``leading'') charged particle, with a further per-event
distinction into ``min'' and ``max'' sides within the transverse region. This
construction produces a set of observables collectively sensitive to each of the
components of the underlying event. The observables in this study
characterise the correlations of a)~each azimuthal region's mean \ptsum and mean
charged-particle multiplicity densities with the transverse momentum, \ptlead, of the
event's leading charged particle; and b)~the per-event mean charged-particle~\pt
with both \ptlead and the transverse azimuthal region's
charged-particle multiplicity. 
The correlations with \ptlead characterise the degree of connection between the
underlying event and the hardest partonic scattering in the $pp$ collision,
while correlations with the transverse regions' charged-particle multiplicities 
probe the particle production mechanisms between underlying-event components.

 \begin{figure*}[t!]
   \begin{center}
     \includegraphics[width=\picwtwores\textwidth]{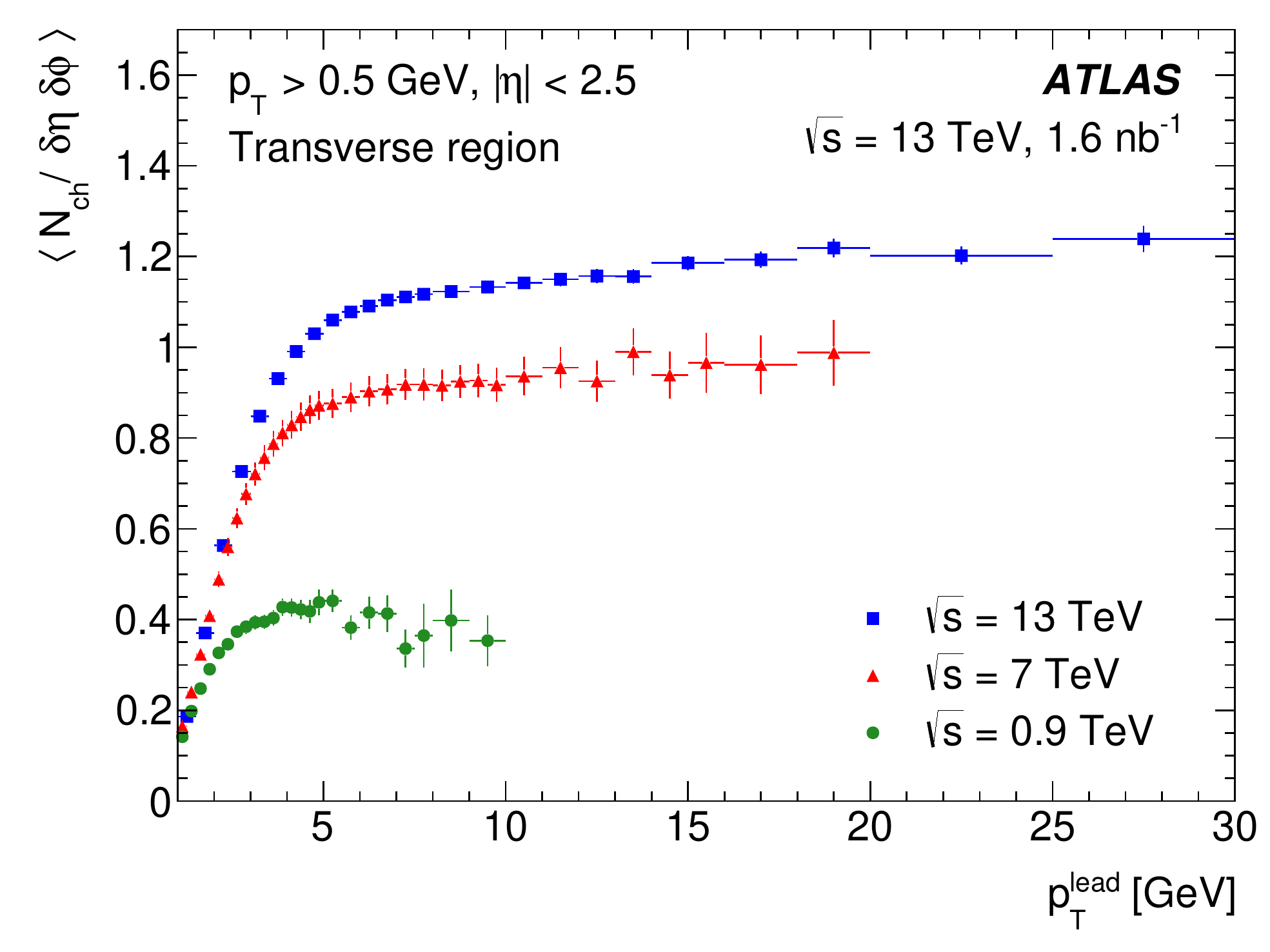}
     \includegraphics[width=\picwtwores\textwidth]{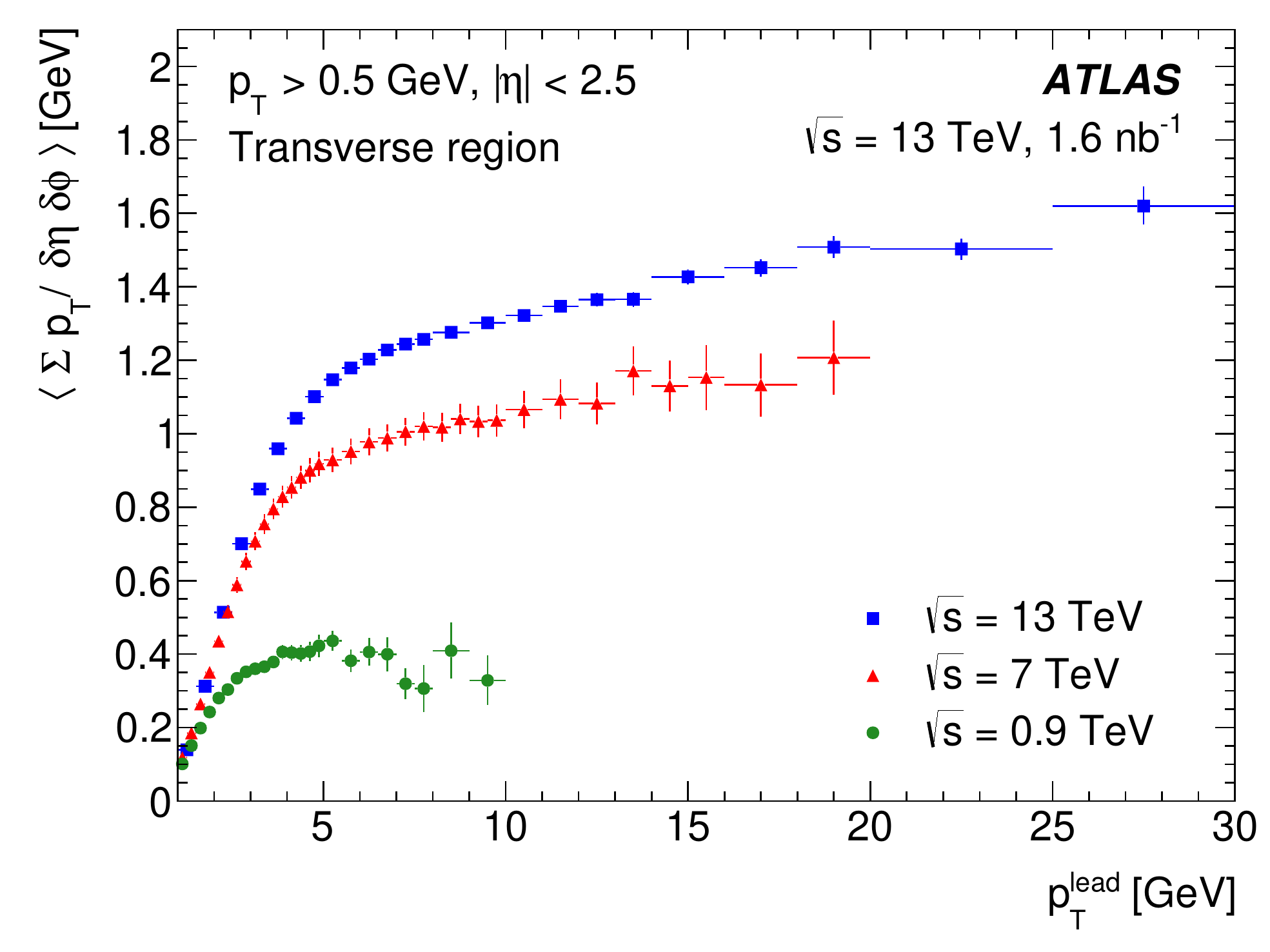}
   \end{center}
   \caption {Mean charged-particle multiplicity (left) and \ptsum (right) densities
     as a function of transverse momentum of the leading charged particle measured for
     $\sqrt{s}= 0.9, \SI{7}{\TeV}$~\cite{Aad:2010fh} and$~\SI{13}{\TeV}$ centre-of-mass energies.
     The fiducial acceptance definitions of the $\sqrt{s} = 0.9$ and
     \SI{7}{\TeV} measurements did not exclude charged strange baryons, but this
     effect is limited to a few percent at most.}
   \label{fig:multieng}
 \end{figure*}

The presented measurement improves upon previous ATLAS measurements of the
underlying event using leading-track alignment, both in the reach in $\ptlead$
and the precision achieved. Figure~\ref{fig:multieng} shows the transverse
region's mean \nch and \ptsum densities in bins of \ptlead as measured in
leading-charged-particle UE studies at
$\sqrt{s}=0.9~\text{and}~\SI{7}{\TeV}$,
compared with the new \SI{13}{\TeV} data. An increase
in UE activity of approximately 20\% is observed when going from \SI{7}{\TeV} to
\SI{13}{\TeV} $pp$ collisions.

These results are a crucial input to physics studies in LHC $pp$ collisions
throughout the \SI{13}{\TeV} run, since the underlying event is
present in all inelastic collision processes. Comparisons against predictions
from several commonly used MC generator configurations indicate that for most
observables the models describe the underlying-event data to better than 5\%
accuracy. But this level of variation is far greater than the experimental
uncertainty of the measurements, and there is evidence of systematic
mismodelling. The model defects are particularly acute in the transition from
generic soft inelastic interactions -- so-called ``minimum bias'' scattering --
to secondary interactions in the presence of hard partonic scattering.  Some
improvement to MC tunes in light of these data will hence be of benefit to
physics studies in LHC Run~2. The \Epos MC generator, specialised for simulation
of inclusive soft QCD processes, displays particularly discrepant features as
the \ptlead scale increases, casting doubt on its suitability for modelling LHC
multiple $pp$ interactions despite currently providing the best description of
minimum-bias data.

This study hence provides important data from the new collider energy frontier
for the tuning of MC generators in LHC Run~2, and further incentives for the
development of a nucleon collision model capable of describing collider data in
both hard and soft scatterings and at multiple centre-of-mass energies.

\section*{Acknowledgements}

We thank CERN for the very successful operation of the LHC, as well as the
support staff from our institutions without whom ATLAS could not be
operated efficiently.

We acknowledge the support of ANPCyT, Argentina; YerPhI, Armenia; ARC, Australia; BMWFW and FWF, Austria; ANAS, Azerbaijan; SSTC, Belarus; CNPq and FAPESP, Brazil; NSERC, NRC and CFI, Canada; CERN; CONICYT, Chile; CAS, MOST and NSFC, China; COLCIENCIAS, Colombia; MSMT CR, MPO CR and VSC CR, Czech Republic; DNRF and DNSRC, Denmark; IN2P3-CNRS, CEA-DSM/IRFU, France; GNSF, Georgia; BMBF, HGF, and MPG, Germany; GSRT, Greece; RGC, Hong Kong SAR, China; ISF, I-CORE and Benoziyo Center, Israel; INFN, Italy; MEXT and JSPS, Japan; CNRST, Morocco; FOM and NWO, Netherlands; RCN, Norway; MNiSW and NCN, Poland; FCT, Portugal; MNE/IFA, Romania; MES of Russia and NRC KI, Russian Federation; JINR; MESTD, Serbia; MSSR, Slovakia; ARRS and MIZ\v{S}, Slovenia; DST/NRF, South Africa; MINECO, Spain; SRC and Wallenberg Foundation, Sweden; SERI, SNSF and Cantons of Bern and Geneva, Switzerland; MOST, Taiwan; TAEK, Turkey; STFC, United Kingdom; DOE and NSF, United States of America. In addition, individual groups and members have received support from BCKDF, the Canada Council, CANARIE, CRC, Compute Canada, FQRNT, and the Ontario Innovation Trust, Canada; EPLANET, ERC, ERDF, FP7, Horizon 2020 and Marie Sk{\l}odowska-Curie Actions, European Union; Investissements d'Avenir Labex and Idex, ANR, R{\'e}gion Auvergne and Fondation Partager le Savoir, France; DFG and AvH Foundation, Germany; Herakleitos, Thales and Aristeia programmes co-financed by EU-ESF and the Greek NSRF; BSF, GIF and Minerva, Israel; BRF, Norway; CERCA Programme Generalitat de Catalunya, Generalitat Valenciana, Spain; the Royal Society and Leverhulme Trust, United Kingdom.

The crucial computing support from all WLCG partners is acknowledged gratefully, in particular from CERN, the ATLAS Tier-1 facilities at TRIUMF (Canada), NDGF (Denmark, Norway, Sweden), CC-IN2P3 (France), KIT/GridKA (Germany), INFN-CNAF (Italy), NL-T1 (Netherlands), PIC (Spain), ASGC (Taiwan), RAL (UK) and BNL (USA), the Tier-2 facilities worldwide and large non-WLCG resource providers. Major contributors of computing resources are listed in Ref.~\cite{ATL-GEN-PUB-2016-002}.

\printbibliography

\newpage 
\begin{flushleft}
{\Large The ATLAS Collaboration}

\bigskip

M.~Aaboud$^\textrm{\scriptsize 137d}$,
G.~Aad$^\textrm{\scriptsize 88}$,
B.~Abbott$^\textrm{\scriptsize 115}$,
J.~Abdallah$^\textrm{\scriptsize 8}$,
O.~Abdinov$^\textrm{\scriptsize 12}$,
B.~Abeloos$^\textrm{\scriptsize 119}$,
S.H.~Abidi$^\textrm{\scriptsize 161}$,
O.S.~AbouZeid$^\textrm{\scriptsize 139}$,
N.L.~Abraham$^\textrm{\scriptsize 151}$,
H.~Abramowicz$^\textrm{\scriptsize 155}$
% ...
\bigskip
\\
% ...
$^{8}$ Department of Physics, The University of Texas at Arlington, Arlington TX, United States of America\\
% ...
$^{12}$ Institute of Physics, Azerbaijan Academy of Sciences, Baku, Azerbaijan\\
% ...
$^{88}$ CPPM, Aix-Marseille Universit{\'e} and CNRS/IN2P3, Marseille, France\\
% ...
$^{115}$ Homer L. Dodge Department of Physics and Astronomy, University of Oklahoma, Norman OK, United States of America\\
% ...
$^{119}$ LAL, Univ. Paris-Sud, CNRS/IN2P3, Universit{\'e} Paris-Saclay, Orsay, France\\
% ...
$^{137}$ $^{(a)}$ Facult{\'e} des Sciences Ain Chock, R{\'e}seau Universitaire de Physique des Hautes Energies - Universit{\'e} Hassan II, Casablanca; $^{(b)}$ Centre National de l'Energie des Sciences Techniques Nucleaires, Rabat; $^{(c)}$ Facult{\'e} des Sciences Semlalia, Universit{\'e} Cadi Ayyad, LPHEA-Marrakech; $^{(d)}$ Facult{\'e} des Sciences, Universit{\'e} Mohamed Premier and LPTPM, Oujda; $^{(e)}$ Facult{\'e} des sciences, Universit{\'e} Mohammed V, Rabat, Morocco\\
% ...
$^{139}$ Santa Cruz Institute for Particle Physics, University of California Santa Cruz, Santa Cruz CA, United States of America\\
% ...
$^{151}$ Department of Physics and Astronomy, University of Sussex, Brighton, United Kingdom\\
% ...
$^{155}$ Raymond and Beverly Sackler School of Physics and Astronomy, Tel Aviv University, Tel Aviv, Israel\\
% ...
$^{161}$ Department of Physics, University of Toronto, Toronto ON, Canada\\
% ...
$^{a}$ Also at Department of Physics, King's College London, London, United Kingdom\\
$^{b}$ Also at Institute of Physics, Azerbaijan Academy of Sciences, Baku, Azerbaijan\\
$^{c}$ Also at Novosibirsk State University, Novosibirsk, Russia\\
$^{d}$ Also at TRIUMF, Vancouver BC, Canada\\
$^{e}$ Also at Department of Physics {\&} Astronomy, University of Louisville, Louisville, KY, United States of America\\
% ...
\end{flushleft}




\end{document}